\documentclass[12pt]{article}
\usepackage{scicite}
\usepackage{graphicx}
\usepackage{times}
\usepackage{amsmath}
\usepackage{gensymb}
\usepackage{amssymb}
\usepackage{xspace}
\usepackage{cases}

\usepackage[mathlines]{lineno}
\modulolinenumbers[2]

\renewcommand{\eqref}[1]{Eq.\,(\ref{#1})}
\newcommand{\figref}[1]{Fig.\,\ref{#1}}
\newcommand{\tableref}[1]{{\bf Table.\,\ref{#1}}}

\newcommand{\mean}[1]{\overline{#1}}

\newcommand{\ec}[0]{$\,\varepsilon(\theta_{\rm c})~$}
\newcommand{\et}[0]{$\,\varepsilon(\theta_{\rm t})~$}
\newcommand{\ea}[0]{$\,\varepsilon(\theta_{\rm a})~$}

\newcommand{\xc}[0]{$\,\tau(\theta_{\rm c})~$}
\newcommand{\xt}[0]{$\,\tau(\theta_{\rm t})~$}

\newcommand{\cflow}[0]{$\theta_{\rm c}$-flow\xspace}
\newcommand{\tflow}[0]{$\theta_{\rm t}$-flow\xspace}
\newcommand{\aflow}[0]{$\theta_{\rm a}$-flow\xspace}

\newcommand{\cflows}[0]{$\theta_{\rm c}$-flows\xspace}
\newcommand{\tflows}[0]{$\theta_{\rm t}$-flows\xspace}
\newcommand{\aflows}[0]{$\theta_{\rm a}$-flows\xspace}

\newcommand{\SSecCoupling}[0]{\textit{\textbf{SM.\,Sec.S1}}\xspace}
\newcommand{\SSecBEM}[0]{\textit{\textbf{SM.\,Sec.S2}}\xspace}
\newcommand{\SSecModel}[0]{\textit{\textbf{SM.\,Sec.S3}}\xspace}
\newcommand{\SSecPtx}[0]{\textit{\textbf{SM.\,Sec.S4}}\xspace}

\title{The younger flagellum coordinates the beating in \textit{C.~reinhardtii}} 

\author
{Da Wei$^{1,3}$,Greta Quaranta$^{2}$, Marie-Eve Aubin-Tam$^{1}\dagger$, Daniel S.W. Tam$^{2}\ast$\\
\\
\normalsize{$^{1}$Department of Bionanoscience, Delft University of Technology,}\\
\normalsize{2628CJ Delft, Netherlands.}\\
\normalsize{$^{2}$Laboratory for Aero and Hydrodynamics, Delft University of Technology,}\\
\normalsize{2628CD Delft, Netherlands.}\\
\normalsize{$^{3}$Beijing National Laboratory for Condensed Matter Physics, Institute of Physics, }\\
\normalsize{Chinese Academy of Sciences; Beijing 100190, China.}\\
 \\
\normalsize{$\dagger$Corresponding author. Email: m.e.aubin-tam@tudelft.nl;}\\
\normalsize{$\ast$Corresponding author. Email: d.s.w.tam@tudelft.nl.}
}

\date{}

\begin{document} 

\baselineskip18pt

\maketitle 

\clearpage
\begin{abstract}
Eukaryotes swim with coordinated flagellar (ciliary) beating and steer by fine-tuning the coordination. The model organism for studying flagellate motility, \emph{C. reinhardtii} (CR), employs synchronous, breast-stroke-like flagellar beating to swim, and it modulates the beating amplitudes differentially to steer. This strategy hinges on both inherent flagellar asymmetries (e.g. different response to chemical messengers) and such asymmetries being effectively coordinated in the synchronous beating. In CR, the synchrony of beating is known to be supported by a mechanical connection between flagella, however, how flagellar asymmetries persist in the synchrony remains elusive. For example, it has been speculated for decades that one flagellum leads the beating, as its dynamic properties (i.e. frequency, waveform, etc.) appear to be copied by the other one. In this study, we combine experiments, computations, and modeling efforts to elucidate the roles played by each flagellum in synchronous beating. With a non-invasive technique to selectively load each flagellum, we show that the coordinated beating essentially responds to only load exerted on the \textit{cis} flagellum; and that such asymmetry in response derives from a unilateral coupling between the two flagella.
Our results highlight a distinct role for each flagellum in coordination and have implication for biflagellates' tactic behaviors.
\end{abstract}

\noindent\textbf{One-Sentence Summary}: The younger flagellum of \emph{C. reinhardtii} coordinates the synchronous beating and couples to external forces.
\clearpage

\section*{Introduction}
The ability to swim towards desirable environments and away from hazardous ones is fundamental to the survival of many microorganisms. These so-called tactic behaviors are exhibited by many motile microorganisms ranging from bacteria~\cite{Berg1972,Smriga2016} to larger flagellates and ciliates~\cite{Hegemann2009,Ueki2010,Stehnach2021}. Different microorganisms have developed specific strategies for steering, depending on the tactic behavior and on their specific sensory and motility repertoire. For example, bacteria modulate the tumbling rate~\cite{Berg1972} while flagellates and ciliates modulate the waveform~\cite{Brokaw1974,Gong2020,Gadelha2020,Bennett2015Interface}, amplitude~\cite{Ruffer1991,Ueki2017} and frequency of their flagellar/ciliary~\cite{Naitoh1972,Ueki2010} beating. The goal of these active modulations of the motility is to achieve a spatially asymmetric generation of propulsive force to steer the organism.  

{\it C. reinhardtii} (CR), the model organism for studies of flagellar motility, achieves tactic navigation by a fine-tuned differential modulation on its two flagella. Studying this organism offers great opportunities to look into how flagella coordinate with each other and how such coordination helps facilitate targeted steering.
CR has a symmetric cell body and two near-identical flagella inherited from the common ancestors of land plants and animals~\cite{Merchant2007}. It swims by beating its two flagella synchronously and is capable of photo- and chemotaxis~\cite{Ruffer1991, Choi2016}. For this biflagellated organism, effective steering hinges on both flagellar asymmetry and flagellar coordination. On the one hand, the two flagella must be asymmetric to respond differentially to stimuli~\cite{Ruffer1990_1,Ruffer1991}; on the other hand, the differential responses must be coordinated by the cell such that the beating would remain synchronized to guarantee effective swimming. Understanding this remarkable feat requires knowledge about both flagellar asymmetry and coordination. 

The two flagella are known to be asymmetric in several, possibly associated, aspects. First of all, they differ in developmental age~\cite{Holmes1989,Dutcher2016}. 
The flagellum closer to the eyespot, the \textit{cis}(-eyespot) flagellum, is always younger than the other one, the \textit{trans}(-eyespot) flagellum. This is because the \textit{cis} is organized by a basal body (BB) that develops from a pre-matured one in the mother cell; and this younger BB also organizes the flagellar root (D4 rootlet) that dictates the eyespot formation~\cite{Mittelmeier2011ChR1onCis}. Second, the two flagella have asymmetric protein composition~\cite{Sakakibara1991, Mackinder2017,Yu2020CAH6}. For example, the \textit{trans} flagellum is richer in CAH6, a protein possibly involved in CO$_2$ sensing~\cite{Choi2016,Mackinder2017}. Finally, the flagella have different dynamic properties~\cite{Kamiya1984,Okita2005, Takada1997}.
Their beating is modulated differentially by second messengers such as calcium~\cite{Kamiya1984,Okita2005} and cAMP~\cite{Saegusa2015}. When beating alone, the \textit{trans} beats at a frequency 30\%-40\% higher than the \textit{cis}~\cite{Kamiya1987,Ruffer1987,Okita2005,Wan2013}; the \textit{trans} also displays an attenuated waveform~\cite{Leptos2013} and a much stronger noise~\cite{Leptos2013,Wan2018}.

Remarkably, despite these inherent asymmetries, CR cells establish robust synchronization between the flagella. 
Such coordination enables efficient swimming and steering of the cells and takes basis on the fibrous connections between flagellar bases~\cite{Quaranta2015,Wan2016}. Intriguingly, in the coordinated beating, both flagella display dynamic properties, i.e., flagellar waveform, beating frequency ($\sim$50 Hz), and frequency fluctuation, that are more similar to those of the \textit{cis} flagellum~\cite{Ruffer1985,Kamiya1987,Leptos2013,Wan2013,Wan2018}. This has led to a long-standing hypothesis that "the \textit{cis} somehow tunes the \textit{trans} flagellum"~\cite{Kamiya1987}. 
This implies that the symmetric flagellar beating ("breast-stroke") observed is the result of interactions between two flagella playing differential roles in coordination. How does the basal coupling make this possible? Recent theoretical efforts show that the basal coupling can give rise to different synchronization modes~\cite{Klindt2017,Liu2018,Guo2021}; and that flagellar dynamics, such as beating frequency, may simply emerge from the interplay between mechanics of basal coupling and bio-activity~\cite{Guo2021}. 
Yet, most theoretical efforts examining flagellar synchronization have assumed two identical flagella, limiting the results' implication for the realistic case. Moreover, little experiments directly probe the flagella’s differential roles during synchronous beating~\cite{Wan2014prl}. Therefore, flagellar coordination in this model organism remains unclear. To clarify the picture experimentally, one needs to selectively force each flagellum, and characterize the dynamics of the flagellar response.

In this study, we address this challenge and devise a non-invasive approach to apply external forces selectively on the \textit{cis}- or the \textit{trans}-flagella. Oscillatory background flows are imposed along an angle with respect to the cell's symmetry axis. Such flows result in controlled hydrodynamic forces, which are markedly different on the two flagella. With experiments, hydrodynamic computations, and modeling, we show definitively that the two flagella are unilaterally coupled, such that the younger flagellum (\textit{cis}) coordinates the beating, whereas the elder one simply copies the dynamic properties of the younger. This also means that only external forces on the \textit{cis} may mechanically fine-tune the coordination. We also study the effect of calcium in the \textit{cis}' leading role as calcium is deeply involved in flagellar asymmetry and hence phototactic steering.
In addition, a well-known mutant that lacks flagellar dominance (\textit{ptx1})~\cite{Horst1993,Okita2005} is examined. Results show that the coordinating role of \textit{cis} does not need environmental free calcium, whereas it does require the genes lost in \textit{ptx1}. Our results discern the differential roles of CR's flagella, highlight an advanced function of the inter-flagellar mechanical coupling, and have implications for biflagellates' tactic motility.

\section*{Experimental scheme for selective loading}

We set out to establish a non-invasive experimental technique that exerts differential loads on the flagella of CR. Following the study by Quaranta et al.~\cite{Quaranta2015}, we induce oscillatory background flows to exert hydrodynamic forcing to flagella of captured cells. With programmed oscillations of the piezoelectric stage, the amplitude, frequency, and direction of the background flows are all controlled, enabling selective loading. 

\begin{figure}[t!]
\centering
\includegraphics[width=0.7\linewidth]{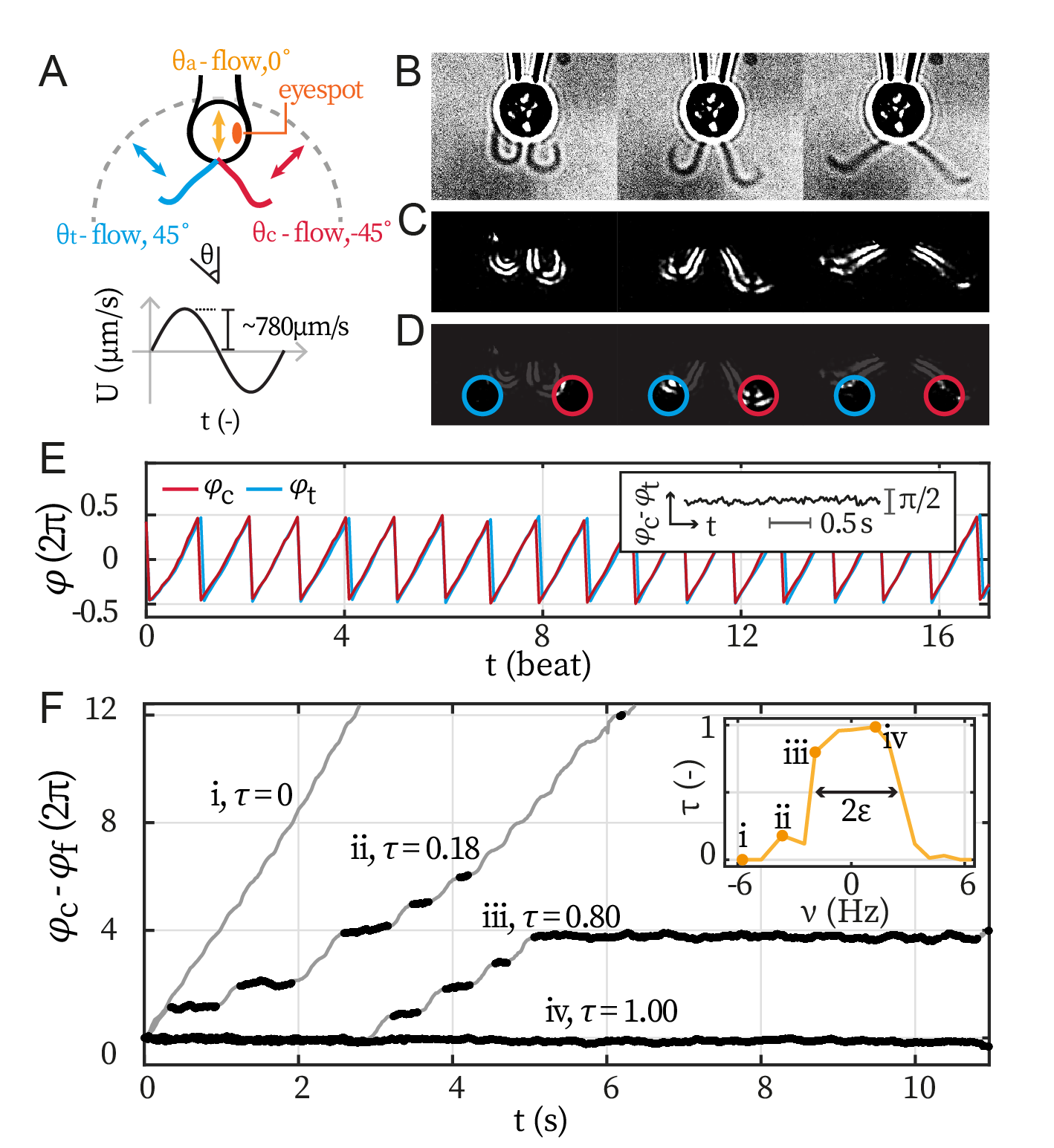}
\vspace{-2mm}
\caption{Experimental workflow. (A) Captured CR cells are subjected to sinusoidal flows of frequency $f_{\rm f}$ along given angles ($\theta$) in the $xy$-plane. Flows along $\theta=-45\degree,0\degree,45\degree$ of same amplitude (780$\pm$50 $\mu$m/s, mean$\pm$std.) are used and termed as shown. (B-E) Extracting flagellar phase $\varphi_{c}$ and $\varphi_{t}$ by image processing. Raw images (B) are thresholded and contrast-adjusted to highlight the flagella (C). Mean pixel values within the user-defined interrogation windows (red and blue circles) capture the raw phases of beating (D), which are then converted to observable-independent phases (E). Inset: phase difference $\varphi_{c} - \varphi_{t}$. (F) Flagella-flow phase dynamics at decreasing detuning $\nu=f_{\rm f}-f_{\rm 0}$ with $f_{\rm 0}$ the cell's beating frequency without external flow. Traces i to iv are taken at detunings marked in the inset. Plateaus marked black represent flow synchronization, whose time fractions $\tau=t_{\rm sync}$/$t_{\rm tot}$ are noted. $t_{\rm tot}$ is the total time of recording. Inset: the flow synchronization profile, $\tau(\nu)$, reports the effective forcing strength 2$\varepsilon$ by its width.}
\label{fig:methods}
\vspace{-3mm}
\end{figure}

To quantitatively estimate the selectivity of the flows along different angles~($\theta$), we compute the flagellar loads under the flows along $\theta=-45\degree$, $0\degree$, and $45\degree$, see \figref{fig:methods}A. Computations based on boundary element methods (BEM) and slender-body theory (SBT) give the real-time drag force $\textbf{F}$ on each flagellum and the power $P$ exerted by the viscous forces on each flagellum. For given realistic flagellar shapes, we compare computed loads with and without external flows. From these we isolate the loads from the induced flows $\textbf{F}_{\rm Flow}$ and $P_{\rm Flow}$ (\textit{Methods}).

\begin{figure}[htb]
\centering
\includegraphics[width=0.7\linewidth]{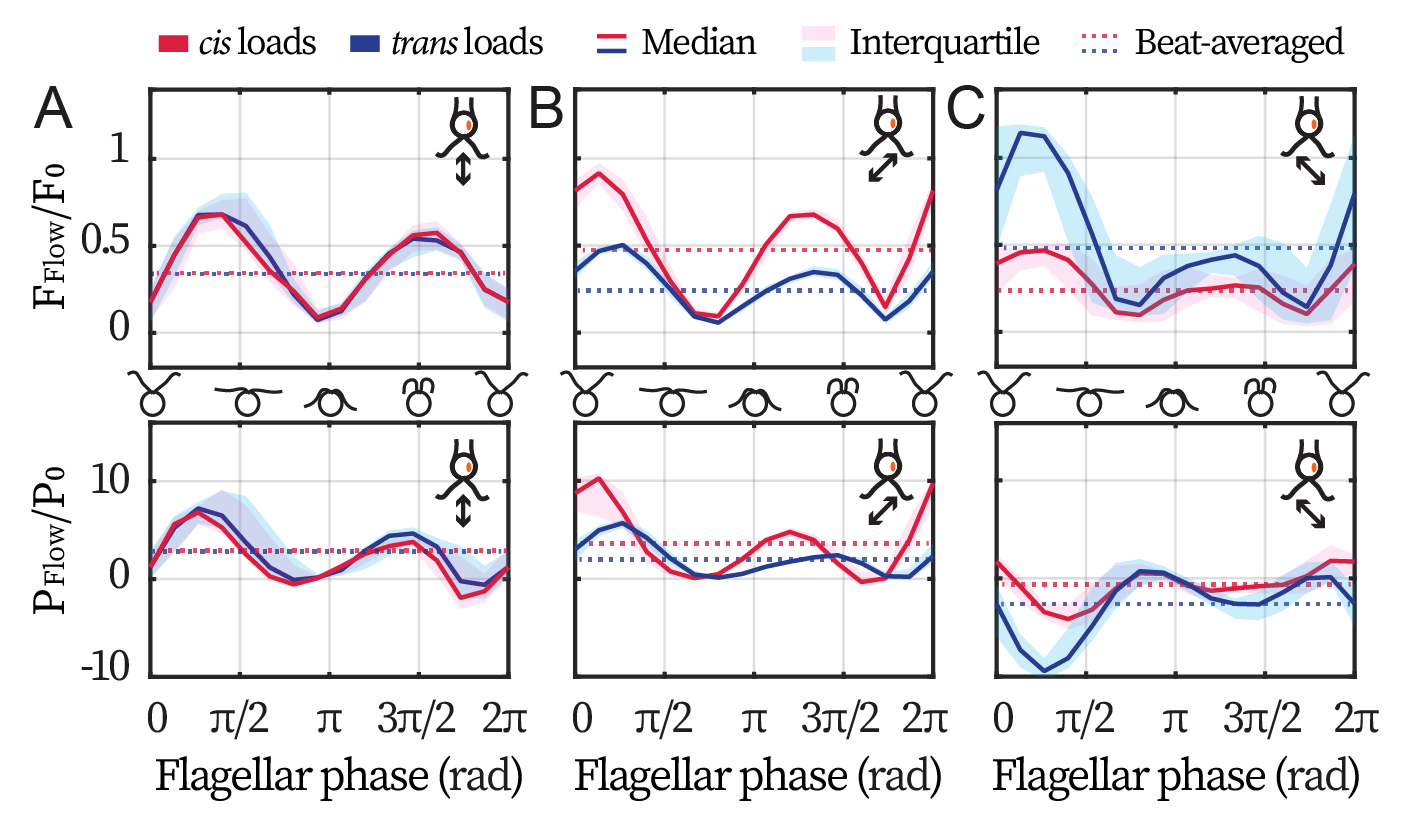}
\vspace{0mm}
\caption{External flagellar loads when beating is synchronized. Force magnitude (upper panels) and power (lower panels) exerted by external flows of $\theta=0\degree$ (A, \aflow), $-45\degree$ (B, \cflow), and $+45\degree$ (C, \tflow). The medians (solid lines) and interquartile ranges (shadings) are computed over $\sim$20 synchronized beats. Dashed horizontal lines: loads averaged over a synchronized beat. 
Force magnitudes and powers are scaled by $F_0$=9.9 pN and $P_0$=1.1 fW respectively. Flagellar phase corresponds to the displayed shapes in the middle $x$-axis.}
\label{fig:BEM_CATFlow}
\end{figure}

Loads on each flagellum under flows of $\theta=0\degree,\ -45\degree,\ 45\degree$ are presented in \figref{fig:BEM_CATFlow}. Upper panels display the magnitude of the drag force $F_{\rm Flow} = |\textbf{F}_{\rm Flow}|$; while lower panels show viscous power $P_{\rm Flow}$. Force magnitudes are scaled by $F_0 = 6\pi\mu R U_0$ = 9.9 pN; while the powers by $P_0 = F_0 U_0= 1.1$ fW. $F_0$ is the Stokes drag on a typical free-swimming cell (radius $R = 5\ \mu$m, speed $U_0=110\ \mu$m/s, water viscosity $\mu= 0.95$ mPa$\cdot$s).

Evidently, along $\theta=0\degree$, flows load the flagella equally (\figref{fig:BEM_CATFlow}A). However, at $\theta=-45\degree$, flows load the \textit{cis} flagellum $\sim2$ times larger than the \textit{trans} (\figref{fig:BEM_CATFlow}B, $F^{\rm c}_{\rm Flow}\approx2F^{\rm t}_{\rm Flow}$); whereas flows at $\theta=45\degree$ do the opposite (\figref{fig:BEM_CATFlow}C). The selectivity also manifests in (the absolute values of) $P_{\rm Flow}$. We do notice that flows along $\theta=+45\degree$ are able to synchronize the flagella with $P_{\rm Flow}<0$, meaning that the flagella are working against the flows, and this shall be discussed in later sections.

Hereon forward, we refer to \cflows, flows for which $\theta=-45 \degree$ and the \textit{cis}-flagellum is selectively loaded. Likewise, \tflows denote flows on $\theta= +45 \degree$ that selectively load the \textit{trans}. \aflows denote the axial flow along $\theta= 0 \degree$.
We next introduce how we quantify the flows' effective forcing strength ($\varepsilon$) on the cell.

Phase dynamics of flagellar beating is extracted from videography~\cite{Quaranta2015, Wei2019, Wei2021}. Recordings are masked and thresholded to highlight the flagella~(\figref{fig:methods}B-C). Then the mean pixel values over time within two sampling windows~(\figref{fig:methods}D) are converted to observable-invariant flagellar phases~\cite{Kralemann2008}, \figref{fig:methods}E. Throughout this study, as \textit{cis} and \textit{trans} always beat synchronously (\figref{fig:methods}E inset), their phases $\varphi_{\rm c,t}$ are used interchangeably as the flagellar phase $\varphi$.
The flagellar phase dynamics under external periodic forcing is described by Adler equation~\cite{Pikovsky2001,Polin2009,Friedrich2016}:
\begin{equation}
    \begin{aligned}
    \frac{d \Delta \varphi }{dt}=-2 \pi \nu - 2 \pi \varepsilon \sin(\Delta \varphi) + \zeta(t).
    \end{aligned}
    \label{adlerEq}
\end{equation}
$\Delta \varphi=\varphi -2 \pi f_{\rm f}t$ is the phase difference between the beating and the forcing, with $f_{\rm f}$ the forcing frequency, and $\varepsilon$ the forcing strength. The detuning $\nu=f_{\rm f}-f_{0}$ is the frequency mismatch between the beating ($f_0$) and forcing. $\zeta(t)$ represents a white noise that satisfies $\langle \zeta (\tau + t) \zeta(\tau) \rangle = 2T_{\rm eff} \delta(t)$, with $T_{\rm eff}$ an effective temperature and $\delta(t)$ the Dirac delta function.

When the forcing strength outweighs the detuning ($\varepsilon>|\nu|$), synchronization with the flow ($d \Delta \varphi /dt=0$) emerges, see the plateaus marked black in \figref{fig:methods}F. We characterize synchronization with $\tau=t_{\rm sync}$/$t_{\rm tot}$, where $t_{\rm sync}$ is the total time of flow synchronization and $t_{\rm tot}$ the flow duration. \figref{fig:methods}F presents the phase dynamics which are representative and range from: no synchronization ($\tau$=0, \textit{i}), unstable synchronization ($0<\tau<1$, \textit{ii-iii}), and stable synchronization ($\tau$=1, \textit{iv}). In this study, the frequency range in $\nu$ for which $\tau\geq0.5$ is used to measure $\varepsilon$ (see \figref{fig:methods}F inset). This method is equivalent to previous fitting-based methods~\cite{Wan2013,Quaranta2015}, see \SSecCoupling. 

\section*{Asymmetric susceptibility to flow synchronization}

\begin{figure}[hbtp]
\vspace{-4mm}
\centering
\includegraphics[width=0.65\linewidth]{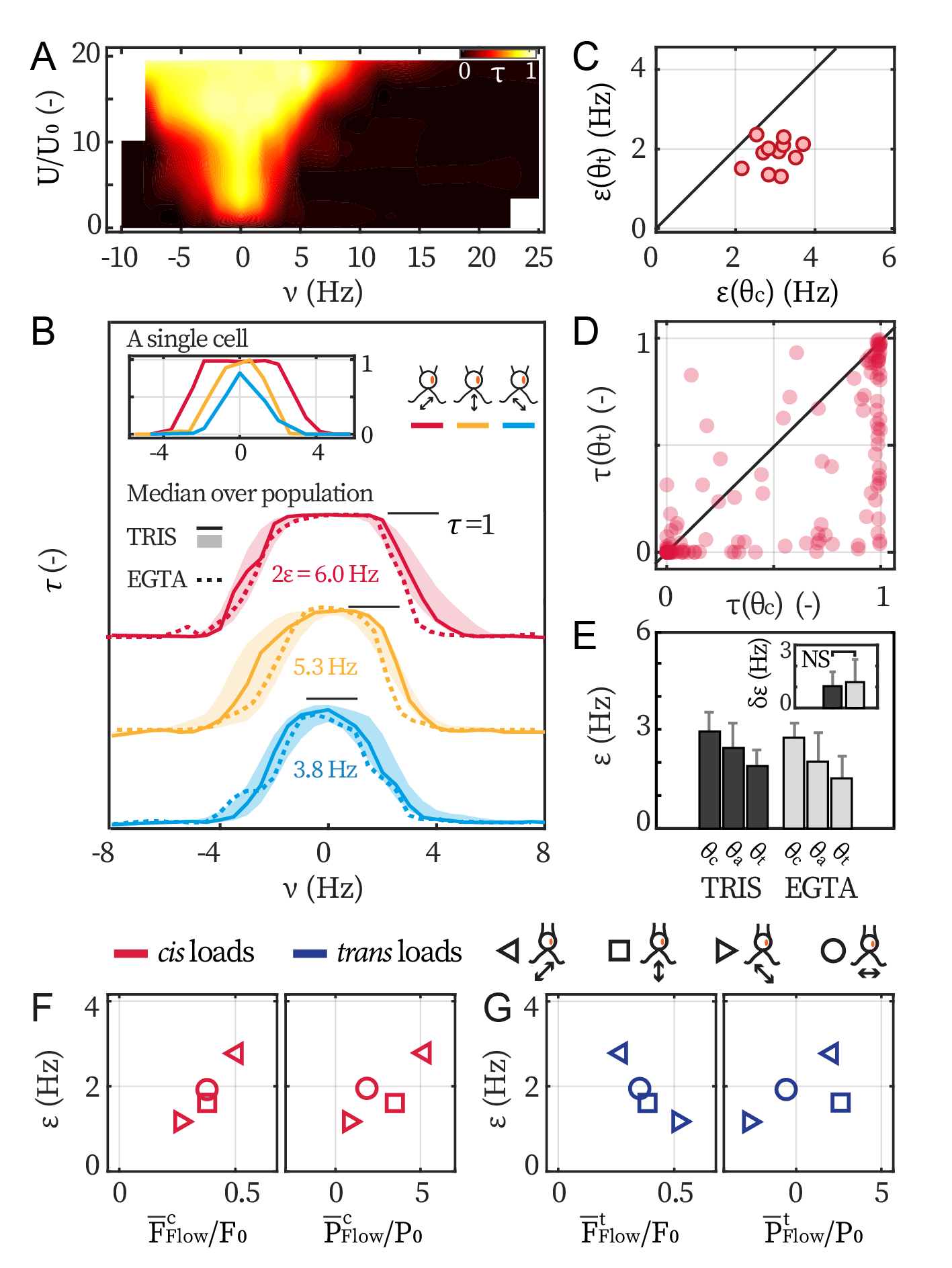}
\caption{Flow synchronization of \textit{wt} cells. (A) Arnold tongue of a representative cell tested with \aflow. The contour is interpolated from N=132 measurements (6 equidistant amplitudes $\times$ 22 equidistant frequencies), and color-coded by the entrained time fraction $\tau$.
(B) The synchronization profiles $\tau(\nu;\theta)$ of a representative {\it wt} cell (inset), the median profile of the TRIS group {\it wt} cells (N=11, solid lines) and the EGTA group (N=6, dashed lines), with either \cflows (red), \aflows (yellow) or \tflows (blue).  Shaded areas are the interquartile ranges for the TRIS group.
(C) Tested {\it wt} cells represented on the \ec$-$\et plane (TRIS group). Solid line: the first bisector line ($y=x$).
(D) Comparing $\tau(\nu;\theta_{\rm c})$ and $\tau(\nu;\theta_{\rm t})$ for each cell at each applied frequency. N=132 pairs of experiments are represented on the \xc$-$\xt plane. More than 90\% of them are below the first bisector line.
(E) The coupling strengths $\varepsilon(\theta)$ of the TRIS group (black) and the EGTA group (gray). Bars and error bars: mean and 1 std., respectively. Inset: $\delta\varepsilon=$\ec$-$\et. NS: not significant, p$>$0.05, Kruskal-Wallis test, One-Way ANOVA. Relations between the forcing strength $\varepsilon$ and the loads on the \textit{cis} (F) and the \textit{trans} flagellum (G).
Markers represent different flow angles, see the drawings.}\label{fig:ct_wt}
\vspace{-3mm}
\end{figure}

Now we examine cell responses to flows of various amplitudes and along different directions. First we explore flow synchronization over a broad range of amplitudes and frequencies. \aflows with frequencies $f_{\rm f} \in [40,\,75]$ Hz and amplitudes $U \in [390, 2340]\ \mu$m/s are imposed. The scanned range covers reported intrinsic frequencies of both the \textit{cis} and \textit{trans} flagellum~\cite{Kamiya1984,Kamiya1987,Ruffer1987,Takada1997}; while the amplitude reaches the maximum instantaneous speed of a beating flagellum ($\sim2000\ \mu$m/s). \figref{fig:ct_wt}A displays the resultant flow-synchronized time fractions $\tau$. Up until the strongest flow amplitude,  the large forces cannot disrupt the synchronized flagellar beating. In addition, synchronization is never established around frequencies other than $f_0$. This shows that the inter-flagella coupling is much stronger than the maximum amplitude of forcing. 

Next we examine the synchronization with the \cflows and \tflows. Flows of a fixed amplitude ($\sim7U_0$) but varying frequencies around $f_0$ are applied to each captured cell (see \textit{Methods}). With these, the flow-synchronized time fraction $\tau$ as a function of the detuning ($\nu$) and flow direction ($\theta_{\rm c,a,t}$) is recorded and helps quantify the flows' effective forcing $\varepsilon(\theta)$. 

Comparing $\tau(\nu;\theta_{\rm c})$ to $\tau(\nu;\theta_{\rm t})$, with $\tau(\nu;\theta_{\rm a})$ as reference, 
we find that \cflows are the most effective in synchronizing the beating (\figref{fig:ct_wt}B). We illustrate this point with the profiles of an exemplary cell (\figref{fig:ct_wt}B inset). First, although both the \cflow (red) and the \tflow (blue) can synchronize the cell at small detunings ($|\nu|<$0.5Hz), the \cflow maintains the synchronization for the whole time (\xc=1), while the \tflow for a slightly smaller time fraction (\xc$\approx$0.85). This is due to phase-slips (step-like changes in $\Delta \varphi(t)$ in \figref{fig:methods}F) between flagella and the flow, and means that the \tflow synchronization is less stable. Additionally, for intermediate detuning (0.5 Hz$<|\nu|<$4 Hz), \xc is always larger than \xt. In some cases, the \cflow synchronizes the cell fully whereas the \tflow fails completely (e.g., at $\nu=-2$ Hz). Together, these results imply that a flow of given amplitude synchronizes flagellar beating more effectively if it selectively loads the \textit{cis}. 

We repeat the experiments with cells from multiple cultures, captured on different pipettes, and with different eyespot orientations ($\sim$50\% heading rightward in the imaging plane) to rule out possible influence from the setup. $\tau(\nu;\theta)$ of N=11 {\it wt} cells tested in the TRIS-minimal medium (pH=7.0) are displayed in \figref{fig:ct_wt}B (labeled as "TRIS"). On average, \ec = 2.9 Hz and is 70\% larger than \et = 1.7 Hz. It bears emphasis that \ec$>$\et holds true for every single cell tested (11/11). In \figref{fig:ct_wt}C, we show this by representing each cell as a point in the \ec-\et plane. A point being below the first bisector line (\ec=\et) indicates that \ec$>$\et for this cell. All cells cluster clearly below the line. This asymmetry manifest equivalently through $\tau$. In \figref{fig:ct_wt}D, each point represents the time fractions of the same cell synchronized by the \cflow and the \tflow at the same frequency. Most points ($>$90\%) are below the first bisector line, meaning that \xc $>$ \xt. 
Altogether, all results show that selectively loading the \textit{cis} flagellum establishes synchronization with the flow more effectively, pointing to \textit{cis} and \textit{trans} playing differential roles in the coordinated beating.

We next study whether this newly observed {\it cis-trans} asymmetry is affected by calcium depletion. Calcium is a critical second messenger for modulating flagellates motility and is deeply involved in phototaxis~\cite{Yoshimura2011}. The depletion of the free environmental calcium is known to degrade flagellar synchronization and exacerbate flagellar asymmetry~\cite{Kamiya1984}. Here we focus on whether calcium depletion affects the asymmetry \ec$>$\et. We deplete environmental calcium by EGTA-chelation, following the protocol in Ref.~\cite{Wakabayashi2009}.
Similar to previous reports~\cite{Kamiya1984,Pazour2005}, the number of freely swimming cells drops significantly in EGTA-containing medium. However, the remaining cells beat synchronously for hours after capture.
For these beating cells, calcium depletion is first confirmed by characterizing their deflagellation behavior. Indeed, calcium depletion is reported to inhibit deflagellation~\cite{Wan2013,Quarmby1994}. In experiments with standard calcium concentration, all cells deflagellated under pipette suction (20/20). For experiments conducted in calcium depleting EGTA-containing medium, we observe deflagellation to occur in none but one cell (1/19). 

After confirming the calcium depletion, we perform the same sets of flow synchronization experiments. The dashed lines in \figref{fig:ct_wt}B show the median synchronization profiles $\tau(\nu;\theta)$ (N=6 cells, labeled as "EGTA"). The flagellar asymmetry is unaffected, see also \figref{fig:ct_wt}E. Note that \ec$>$\et again applies for every single cell tested. The mean values of $\varepsilon$ drop slightly. However, the different effectiveness between \cflows and \tflows, \ec$-$\et, is not affected, see \figref{fig:ct_wt}E inset. 

Finally, we determine how the forcing strength of the flow depends on the hydrodynamic forces exerted by the flow on the flagella. We compute the hydrodynamic beat-averaged loads, $\mean{F}_{\rm Flow} = \int_0 ^{2\pi} F_{\rm Flow} d\varphi /2\pi$, $\mean{P}_{\rm Flow} = \int_0 ^{2\pi} {P}_{\rm Flow} d\varphi /2\pi$, induced by the flow on the \textit{trans} and on the \textit{cis} flagella, see the horizontal lines in \figref{fig:BEM_CATFlow}. These loads are computed for the \cflow, \tflow, \aflow and we also include experiments and computations performed with flows along $\theta=90\degree$ (circles), see \SSecBEM. \figref{fig:ct_wt}F and G represent $\varepsilon$ as a function of the loads on the \textit{cis} and \textit{trans} flagellum respectively, with each symbol representing one of the four different flow directions, see the drawings. We find that the effective forcing strength scales with the time-averaged drag on the \textit{cis}, $\varepsilon \sim \mean{F}^{\rm c}_{\rm Flow}$, while we find no such correlation between $\varepsilon$ and $\mean{F}^{\rm t}_{\rm Flow}$. The linear relation between $\varepsilon$ and $\mean{F}^{\rm c}_{\rm Flow}$ has an intercept near zero ($\varepsilon|_{\mean{F}^{\rm c}_{\rm Flow}=0}\approx0$). Given the total forces on both flagella ($\mean{F}^{\rm c}_{\rm Flow} + \mean{F}^{\rm t}_{\rm Flow}$) for these flows remains almost constant (0.74-0.79$F_0$), the zero-intercept implies that for a hypothetical flow that exerts no load on the \textit{cis} but solely forces the \textit{trans}, it will not be able to synchronize the cell at all. This suggests a negligible contribution of the forcing on the \textit{trans} in establishing synchronization with flows.

\section*{The asymmetry is lost in \emph{ptx1} mutants}
\label{subsec:ptx1}

Furthermore, we examine the flagellar dominance mutant {\it ptx1}. In this mutant, both flagella respond similarly to changes of calcium concentrations~\cite{Horst1993} and have similar beating frequencies when demembranated and reactivated~\cite{Okita2005}. 

\begin{figure}[hp]
\centering
\includegraphics[width=0.7\linewidth]{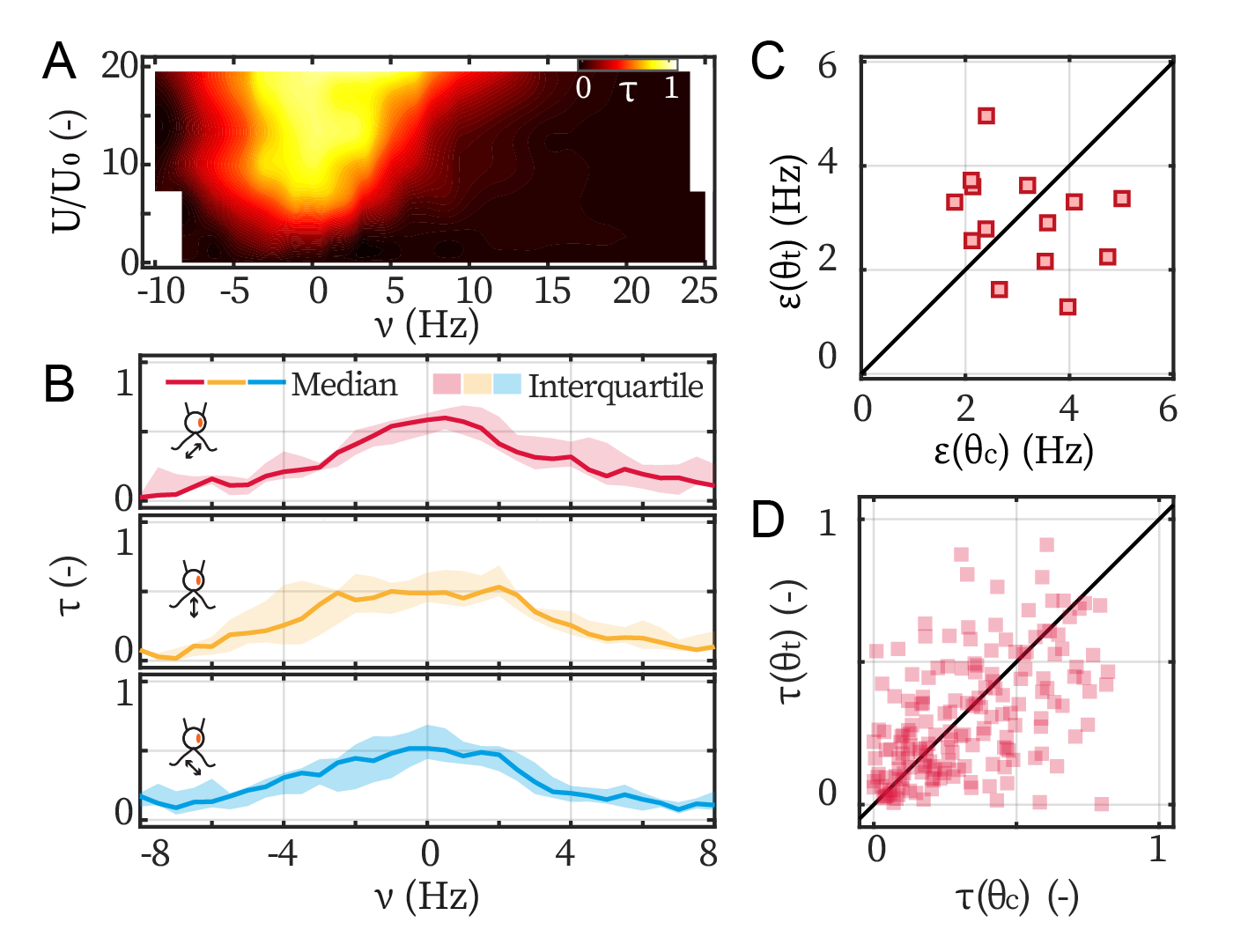}
\vspace{-4mm}
\caption{The asymmetric susceptibility to flow synchronization is lost in the flagellar dominance mutant \emph{ptx1}. 
(A) Arnold tongue of a representative {\it ptx1} cell tested with \aflow. The contour is interpolated from N=132 measurements (6 equidistant amplitudes $\times$ 22 equidistant frequencies). Color bar: the entrained time fraction $\tau=t_{\rm sync}/t_{\rm IP}$.
(B) Flow synchronization profiles $\tau(\nu;\theta)$ of N=14 {\it ptx1} cells, tested with \cflows (red), \aflows (yellow) and \tflows (blue).
(C) \ec and \et of the tested cells. The first bisector line (solid): $y=x$. 
(D) $\tau(\nu;\theta_{\rm c,t})$ for each cell at each applied frequency. N=154 points are present.}
\label{fig:ct_ptx1}
\vspace{-1mm}
\end{figure}

{\it Ptx1} mutants have two modes of coordinated beating, namely, the in-phase (IP) synchronization and the anti-phase (AP) synchronization~\cite{Ruffer1998,Leptos2013}. First, we apply \aflow in the same frequency and amplitude ranges as for {\it wt} cells. We find that the IP mode around $f_0\approx50$ Hz is the only mode that can be synchronized by external flows. We focus on this mode and report $\tau$ as $\tau=t_{\rm sync}/t_{\rm IP}$ for this mutant, where $t_{\rm IP}$ is the total time of IP-beating under the applied flows, see \figref{fig:ct_ptx1}A. Synchronization profiles $\tau(\nu;\theta)$ of {\it ptx1} are shown in \figref{fig:ct_ptx1}B. The median profiles are of similar width and height, indistinguishable from each other, and hence indicate a loss of asymmetric susceptibility to flow synchronization. The loss is further confirmed by the extracted $\varepsilon(\theta)$~\cite{Quaranta2015} and $\tau(\theta)$ (\figref{fig:ct_ptx1}C-D). Cells and synchronization attempts are distributed evenly across the first bisector lines (7/14 cells are below \ec=\et in \figref{fig:ct_ptx1}C, and $\sim$50\% points are below \xc$=$\xt in \figref{fig:ct_ptx1}D). Altogether, all results show consistently that the asymmetry is lost in {\it ptx1}.

\section*{Modeling}
\subsection*{Framework}
To investigate the implications of our experimental results on the coupling between flagella and their dynamics, we develop a model for the system (\SSecModel), representing flagella and external flows as oscillators with directional couplings:
\begin{equation}\label{eq:3Body}
\begin{cases}
    \dot{\varphi_{\rm f}} = 2 \pi f_{\rm f}\\
    \dot{\varphi_{\rm c}} = 2 \pi[ f_{\rm c} 
                - \lambda_{\rm t} \sin(\varphi_{\rm c} \text{-} \varphi_{\rm t}) 
                - \varepsilon_{\rm c} \sin(\varphi_{\rm c} \text{-} \varphi_{\rm f})]
                + \zeta_{\rm c}(t)\\
    \dot{\varphi_{\rm t}} = 2 \pi [f_{\rm t} 
                - \lambda_{\rm c} \sin(\varphi_{\rm t} \text{-} \varphi_{\rm c}) 
                - \varepsilon_{\rm t} \sin(\varphi_{\rm t} \text{-} \varphi_{\rm f})]
                + \zeta_{\rm t}(t).
\end{cases}
\end{equation}
$\varphi_{\rm f,c,t}(t)$ respectively represent the phase of the flow, the \textit{cis}, and the \textit{trans} flagellum. $f_{\rm f,c,t}$ represents the inherent frequency of the forcing (flow), the \textit{cis}, and the \textit{trans} respectively. The phase dynamics of each flagellum is modulated by its interactions with the other flagellum as well as the background flow.
Take the \textit{cis} ($\dot{\varphi}_{\rm c}$) for example, the effect of the \textit{trans} and the forcing on the \textit{cis} are respectively accounted for by the $\lambda_{\rm t}$-term and the $\varepsilon_{\rm c}$-term, see \eqref{eq:3Body}. 
In other words, $\lambda_{\rm t}$ and $\varepsilon_{\rm c}$ measure the sensitivity of the actual \textit{cis}-frequency to the phase differences between oscillators ($\varphi_{\rm c} - \varphi_{\rm t,f}$), see the arrows in \figref{fig:model}A.
Lastly, $\zeta_{\rm c,t}$ represent the white noise of the \textit{cis} and \textit{trans} flagellum respectively. In the following parts, without loss of generality, the noise are assumed equally strong and uncorrelated ($\langle \zeta^2_{\rm c} \rangle = \langle \zeta^2_{\rm t} \rangle$, or $T^{\rm c}_{\rm eff}=T^{\rm t}_{\rm eff}$). Nuanced phase dynamics under differential noise levels can be found in \SSecPtx.

\eqref{eq:3Body} can be readily reduced to \eqref{adlerEq}, which allows us to write the experimentally measured values ($f_0,\ \varepsilon(\theta),\ T_{\rm eff}$) analytically with $\varepsilon_{\rm c,t}$, $\lambda_{\rm c,t}$, and $\zeta_{\rm c,t}$. The asymptotic behavior of the model under the condition $\varphi_{c}\approx\varphi_{t}\approx\varphi_{f}$ are (\SSecModel):
\begin{equation}\label{eq:analytical}
\begin{cases}
        f_0 &= \alpha f_{\rm c} + (1-\alpha)f_{\rm t},\\
        T_{\rm eff} &= \alpha^2 T_{\rm eff}^{\rm c} + (1-\alpha)^2T_{\rm eff}^{\rm t},\\
        \varepsilon(\theta) &= \alpha\varepsilon_{\rm c}(\theta) + (1-\alpha)\varepsilon_{\rm t}(\theta),
\end{cases}
\end{equation}
with $\alpha=\lambda_{\rm c}/(\lambda_{\rm c}+\lambda_{\rm t})$ representing the dominance of \textit{cis}. It is then clear that when $\alpha\approx1$, the coordinated beating will display dynamic properties of the \textit{cis} flagellum.

\figref{fig:model}A illustrates an exemplary modeling scheme describing flagellar beating subjected to \cflows. The direction and thickness of arrows represent coupling direction and strength respectively. The selective loading on the \textit{cis} is represented by $\varepsilon_{\rm c}>\varepsilon_{\rm t}$; while $\lambda_{\rm c} > \lambda_{\rm t}$ reflects that the \textit{cis} has a more dominant role in the coordinated beating. We run Monte-Carlo simulation with \eqref{eq:3Body} using customized \textsc{matlab} scripts.

\subsection*{Coordinated beating under symmetric forcing}
We first model the flow synchronization induced by \aflow (symmetric flagellar loads). In this case, $\varepsilon(\theta) = \alpha\varepsilon_{\rm c}(\theta) + (1-\alpha)\varepsilon_{\rm t}(\theta)$ (\eqref{eq:analytical}) reduces to $\varepsilon = \varepsilon_{\rm c,t}$ and is independent of $\alpha$. We set $\varepsilon_{\rm c,t}$ as 2.4~Hz to match the measured \ea=2.4 Hz (\figref{fig:ct_wt}B).

At similar detunings as in the experimental results in \figref{fig:methods}F, our Monte-Carlo simulations reproduces the phase dynamics with: (i) no flow synchronization, (ii-iii) unstable synchronization, and (iv) stable synchronization (\figref{fig:model}B). Repeating the simulations for varying forcing strength $\varepsilon$ ($=\varepsilon_{\rm c,t}$) and frequency $f_{\rm f}$ yields Arnold tongue diagrams in agreement with those reported from our experiments. The Arnold Tongue for \textit{wt} in \figref{fig:ct_wt}A and \textit{ptx1} in \figref{fig:ct_ptx1}A are reproduced with simulations shown in \figref{fig:model}C and D respectively. The only parameter value changed between \figref{fig:model}C and D is the level of noise ($T^{\rm c,t}_{\rm eff}$), which is increased by an order of magnitude. The differences in phase dynamics between \textit{wt} and \textit{ptx1}, when subjected to symmetric external loading, are therefore accounted by solely varying the noise.

\subsection*{Coordinated beating under selective loading}
\begin{figure}[htbp]
\vspace{-3mm}
\centering
\includegraphics[width=0.59\linewidth]{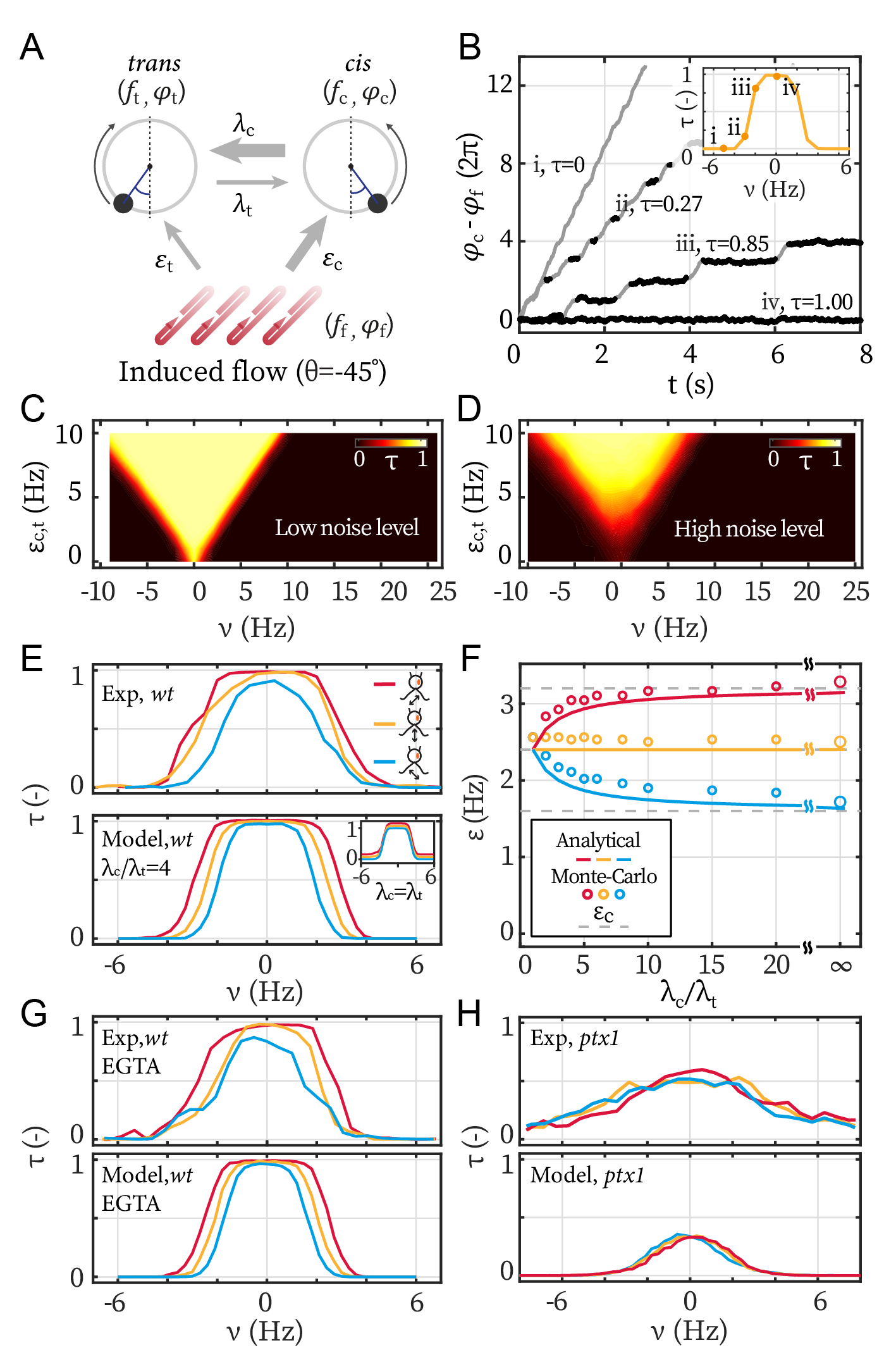}
\vspace{-3mm}
\caption{Modeling the asymmetric flow synchronization. (A) Modeling scheme describing a cell beating under directional flow (\cflow as an example). Arrows represent the directional coupling coefficients with line thickness representing the relative strength. For example, $\lambda_c$ points from \textit{cis} to \textit{trans}, representing how the latter ($\varphi_c$) is sensitive to the former ($\varphi_t$); meanwhile, the arrow of $\lambda_c$ being thicker than $\lambda_t$ means that $\varphi_t$ is much more sensitive to $\varphi_c$ than the other way around.
(B) Modeled phase dynamics of flow synchronization under \aflows, analogous to \figref{fig:methods}F. Reproducing the Arnold tongue diagrams at the noise level of {\it wt}~(C) and {\it ptx1}~(D), analogous to \figref{fig:ct_wt}A and \figref{fig:ct_ptx1}A respectively. (E) Flow synchronization profiles $\tau(\nu;\theta)$ obtained experimentally (upper panel) and by modeling (lower panel). Inset: the modeling results with symmetric inter-flagellar coupling. (F) Effective forcing strength $\varepsilon(\theta)$ as a function of the inter-flagellar coupling asymmetry $\lambda_{\rm c}/\lambda_{\rm t}$. Points: measured from simulation; lines: analytical approximation (\eqref{eq:analytical}); dashed lines: $\varepsilon_{\rm c}$ respectively for the \cflow, \aflow, and \tflow (from top to bottom). (G) Reproducing the flow synchronization of {\it wt} cells under calcium depletion (H) Reproducing results of {\it ptx1}. See \tableref{table:parameters} for the modeling parameters.}
\label{fig:model}
\vspace{-3mm}
\end{figure}

We next model flow synchronization by the \cflows and the \tflows. The selective forcing ($\varepsilon_{\rm c}\neq\varepsilon_{\rm t}$) allows the effect of flagellar dominance ($\lambda_{\rm c}\neq\lambda_{\rm t}$) to manifest in the effective forcing strength $\varepsilon(\theta)$ and hence in the synchronization profiles $\tau(\nu;\theta)$, \figref{fig:model}E. 
Similar to our experimental observations, \cflow synchronizes the coordinated beating over the broadest range of $\nu$ (i.e. largest $\varepsilon$). This is directly attributed to the dominance $\lambda_{\rm c}>\lambda_{\rm t}$: by setting $\lambda_{\rm c}=\lambda_{\rm t}$, the differences between \xc and \xt disappear even under selective loading (\figref{fig:model}E inset). 
\figref{fig:model}F details how the asymmetry of inter-flagellar coupling ($\lambda_{\rm c}/\lambda_{\rm t}$) affects the asymmetry between \xc and \xt. The open symbols represent $\varepsilon(\theta)$ measured from modeled $\tau(\nu;\theta)$ and the lines represent \eqref{eq:analytical}. The difference between \ec and \et increases with $\lambda_{\rm c}/\lambda_{\rm t}$, and they each saturates to reflect only the forcing on the \textit{cis} ($\varepsilon_{\rm c}$, the grey dashed lines). With $f_c=45$~Hz, $f_t=65$~Hz~\cite{Kamiya1987,Okita2005}, and $f_0\approx50$~Hz, we deduce from \eqref{eq:analytical} that $\lambda_{\rm c}=4\lambda_{\rm t}$ for \textit{wt} cells. For {\it wt} cells under calcium depletion, experimental results are reproduced with a lower total forcing strength (\figref{fig:model}G). $\varepsilon_{\rm c} + \varepsilon_{\rm t}$ is set to 4.08~Hz (15\% lower) to reflect the $7\%-20\%$ decrease in $\varepsilon(\theta)$ induced by calcium depletion. 

The {\it ptx1} results are reproduced with a stronger noise ($T^{\rm c,t}_{\rm eff}=9.42\ {\rm rad}^2/{\rm s}$) and a symmetric inter-flagellar coupling $\lambda_{\rm c}/\lambda_{\rm t}=1$, see \figref{fig:model}H and \tableref{table:parameters}. Both changes are necessary for reproducing the synchronization profiles of {\it ptx1} in \figref{fig:model}H: while the stronger noise lowers the maximal values of $\tau(\theta,\nu)$, setting $\lambda_{\rm c}/\lambda_{\rm t}=4$ would still result in \xc$>$\xt in the central range ($|\nu|\lesssim2.4$~Hz). Finally, it is noteworthy that the noise in \textit{ptx1} increases not only because a higher noise value for individual flagella, but also because the \textit{cis}-\textit{trans} coupling has become symmetric. As shown by \eqref{eq:analytical}, the unilateral coupling promotes not only the \textit{cis}-frequency in the synchrony but also the \textit{cis}-noise. Given $T^{\rm c}_{\rm eff} \ll T^{\rm t}_{\rm eff}$ and $\lambda_{\rm c}=4\lambda_{\rm t}$, we confirm with simulations that the \textit{cis} stabilizes the beating frequency of the \textit{trans} and decreases its beating noise. The simulations are in good agreement with experimental noise measurements, see \SSecPtx for details.

\section*{Discussion}
The two flagella of {\it C. reinhardtii} have long been known to have inherently different dynamic properties such as frequency, waveform, level of active noise, and responses to second messengers~\cite{Kamiya1987,Okita2005,Leptos2013,Wan2018,Saegusa2015}. Intriguingly, when connected by basal fibers and beating synchronously, they both adopt the kinematics of the \textit{cis}-(eyespot) flagellum, which led to the assumption that the flagella may have differential roles in coordination. In this work, we test this hypothesis by employing oscillatory flows applied from an angle with respect to the cells' symmetry axis and thus exert biased loads on one flagellum.

Without an exception, in {\it wt} cells, \cflows, the ones that selectively load the \textit{cis} flagellum, are always more effective in synchronizing the flagellar beating than the \tflows. This is shown by the larger effective forcing strengths (\ec$>$\et, \figref{fig:ct_wt}B-C) and larger synchronized time fractions (\xc$>$\xt, \figref{fig:ct_wt}D). Mapping the measured forcing strength $\varepsilon(\theta)$ as a function of the loads, we find empirically that $\varepsilon \propto \mean{F}_{\rm Flow}^{\rm c}$~(\figref{fig:ct_wt}F) and that \textit{trans}-loads appear to matter negligibly. These observations all indicate that the \textit{cis}-loads determine whether an external forcing can synchronize the cell. Moreover, this point is further highlighted by an unexpected finding: when \tflows are applied, the \textit{trans} flagellum always beats against the external flow ($P_{\rm Flow}^{\rm t}<0$) and the only stabilizing factor for flow synchronization is the \textit{cis} flagellum working along with the flow during the recovery stroke (\figref{fig:BEM_CATFlow}C lower panel
). These observations definitively prove that the two flagella have differential roles in the coordination and interestingly imply that flagella are coupled to external flow only through the \textit{cis}.

To have a mechanistic understanding of this finding, we model the system with \eqref{eq:3Body}. 
In the model, selective hydrodynamic loading and flagellar dominance in the coordinated beating are respectively represented by $\varepsilon_{\rm c} \neq \varepsilon_{\rm t}$ and $\lambda_{\rm c} \neq \lambda_{\rm t}$.
Setting out from the model, we obtain closed-form expressions for observables such as $f_0$ and $\varepsilon$ (\eqref{eq:analytical}), which illustrate how flagellar dominance and selective loading affect the coordinated flagellar beating. Moreover, with Monte-Carlo simulation, we clarified the interplay between flows and flagella (\SSecModel), and reproduces all experimental observations. 

With the model, we show that a "dominance" of the \textit{cis} ($\lambda_{\rm c}>\lambda_{\rm t}$) is sufficient to explain why the coordinated flagellar beating bears the frequency and the noise level of the \textit{cis} flagellum. In the model, such dominance means that the \textit{cis}-phase is much less sensitive to the \textit{trans}-phase than the other way around. We then reproduce the phase dynamics of flow synchronization at varying detunings (\figref{fig:model}B), amplitudes (\figref{fig:model}C), and noise (\figref{fig:model}D). Exploiting the observation that the coordination between flagella cannot be broken by external flows up to the strongest ones tested ($\varepsilon^{\rm max}\sim10$~Hz, \figref{fig:ct_wt}A), we quantify the lower limit of the total basal coupling, $\lambda_{\rm c}+\lambda_{\rm t}$, to be approximately 40~Hz (deduced in \SSecModel), which is an order magnitude larger than the hydrodynamic inter-flagellar coupling~\cite{Quaranta2015, Brumley2014, Klindt2016, Pellicciota2020hydrosyncCilia}.

The modulation of flagellar dominance mediates tactic behaviors~\cite{Kamiya1984, Horst1993, Pazour2005, Okita2005}. Calcium is hypothesized to be underlying the modulation of dominance, as it causes the connecting fiber between flagella to contract~\cite{Hayashi1998}, modulates the \textit{cis-} and \textit{trans} activity (e.g. beating amplitude) differentially~\cite{Kamiya1984}, and calcium influx comprises the initial step of CR's photo-~\cite{Harz1991} and mechanoresponses~\cite{Yoshimura2011}.
We therefore investigate flagellar coupling in the context of tactic steering by depleting the environmental free calcium and hence inhibiting signals of calcium influxes.  
Cells are first acclimated to calcium depletion, and then tested with the directional flows. 
Our results show that the \textit{cis} dominance does not require the involvement of free environmental calcium. Calcium depletion merely induces an overall drop in the forcing strength perceived by the cell $\varepsilon(\theta)$ ($7\%-20\%$), which is captured by reducing $\varepsilon_{\rm c}+\varepsilon_{\rm t}$ for 15\% (mean drop) in the model (\figref{fig:model}G). 
Together, our results indicate that the leading role of \textit{cis}, is an inherent property, that does not require active influx of external calcium, and possibly reflects an intrinsic mechanical asymmetry of the cellular mesh that anchors the two flagella into the cell body.

In {\it ptx1} cells, a lack of flagellar dominance ($\lambda_{\rm c}=\lambda_{\rm t}$) and a stronger noise level help reproduce our experimental observations. 
Previous studies suggested that both flagella of {\it ptx1} are similar to the wildtype \textit{trans}~\cite{Okita2005}, and that the noise levels of this mutant's synchronous beating are much greater than those of \textit{wt}~\cite{Leptos2013} (see also \SSecPtx). If both flagella and their anchoring roots indeed have the composition of the wildtype \textit{trans}, such symmetry would predict $\lambda_{\rm c}=\lambda_{\rm t}$. This symmetric coupling renders the noise of \textit{ptx1} $T_{\rm eff} = T_{\rm eff}^{\rm t}$ (\eqref{eq:analytical}), which is about an order of magnitude larger than the noise of \textit{wt} $T_{\rm eff} \approx T_{\rm eff}^{\rm c}$.

The comparison between \textit{ptx1} and \textit{wt} highlights an intriguing advantage of the observed unilateral coupling ($\lambda_{\rm c}\gg\lambda_{\rm t}$); that is, it strongly suppresses the high noise of the \textit{trans}. Considering that the \textit{trans} is richer in CAH6 protein and this protein's possible role in inorganic carbon sensing\cite{Mackinder2017,Choi2016}, the potential sensing role of the \textit{trans} is worth noticing. Assuming the strong noise present in the \textit{trans} originates from the biochemical processes related to sensing, then the unilateral coupling effectively prevents such noise from perturbing the cell's synchronous beating and effective swimming. In this way, the asymmetric coupling may combine the benefit of having a stable \textit{cis} as the driver while equipping a noisy \textit{trans} as a sensor.


\section*{Material and methods}
\subsection*{Cell culture}
CR wildtype ({\it wt}) strain cc125 (mt+) and flagellar dominance mutant {\it ptx1} cc2894 (mt+) are cultured in TRIS-minimal medium (pH=7.0) with sterile air bubbling, in a 14h/10h day-night cycle. Experiments are performed on the 4th day after inoculating the liquid culture, when the culture is still in the exponential growth phase and has a concentration of $\sim 2 \times 10^5 $ cells/ml. Before experiments, cells are collected and resuspended in fresh TRIS-minimal (pH=7.0).

\subsection*{Calcium depletion}
In calcium depletion assays, cells are cultured in the same fashion as mentioned above but washed and resuspended in fresh TRIS-minimal medium + 0.5 mM EGTA (pH=7.0).
Free calcium concentration is estimated to drop from 0.33 mM in the TRIS-minimal medium, to 0.01 \textmu{M} in the altered  medium~\cite{Wakabayashi2009}. Experiments start at least one hour after the resuspension in order to acclimate the cells.

\subsection*{Experimental setup}
\label{sec:setup}
Single cells of CR are studied following a protocol similar to the one described in~\cite{Quaranta2015}. Cell suspensions are filled into a customized flow chamber with an opening on one side. The air-water interface on that side is pinned on all edges and is sealed with silicone oil. A micropipette held by micromanipulator (SYS-HS6, WPI) enters the chamber and captures single cells by aspiration. The manipulator and the captured cell remain stationary in the lab frame of reference, while the flow chamber and the fluid therein are oscillated by a piezoelectric stage (Nano-Drive, Mad City Labs), such that  external flows are applied to the cell. Frequencies and amplitudes of the oscillations are individually calibrated by tracking micro-beads in the chamber. Bright field microscopy is performed on an inverted microscope (Nikon Eclipse Ti-U, 60$\times$ water immersion objective). Videos are recorded with a sCMOS camera (LaVision PCO.edge) at 600-1000 Hz. 

\subsection*{Measurement scheme}
The flagellar beating of each tested cell is recorded before, during, and after the application of the flows. We measure the cell's average beating frequency $f_0$ over 2 s ($\sim$100 beats). For {\it ptx1} cells, $f_0$ is reported for the in-phase (IP) synchronous beating. Unless otherwise stated, directional flows ($\theta=0,\pm45\degree$) are of the same amplitude (780$\pm$50 $\mu$m/s, mean$\pm$std), similar to those used in Ref.\cite{Quaranta2015}. Flow frequencies $f_{\rm f}$ are scanned over [$f_0 - 7$, $f_0 + 7$] Hz for each group of directional flows.  

\subsection*{Computation of the flagellar loads}
\label{sec:BEM}
To quantify the hydrodynamic forces on the flagella, we first track realistic flagellar deformation from videos wherein background flows are applied. Then we employ a hybrid method combining boundary element method (BEM) and slender-body theory~\cite{Keller1976, Wei2021} to compute the drag forces exerted on each flagellum and the forces' rates of work. In this approach, each flagellum is represented as a slender-body~\cite{Keller1976} with 26 discrete points along its centerline and the time-dependent velocity of each of the 26 points is calculated by its displacement across frames. The cell body and the pipette used to capture the cell are represented as one entity with a completed double layer boundary integral equation~\cite{power1987second}. Stresslet are distributed on cell-pipette's surface; while stokeslet and rotlet of the completion flow are distributed along cell-pipette's centerline~\cite{Keaveny2011}. The no-slip boundary condition on the cell-pipette surface is satisfied at collocation points. Lastly, stokeslets are distributed along the centerlines of the flagella, so that no-slip boundary conditions are met on their surfaces. Integrating the distribution of stokeslets $\textbf{f}(s)$ over a flagellar shape, one obtains the total drag force $\textbf{F}=\int \textbf{f}(s) ds$ is obtained. Similarly, the force's rate of work is computed as $P= \int\textbf{f}(s) \cdot \textbf{U}(s) ds$, where $\textbf{U}(s)$ is the velocity of the flagellum at the position $s$ along the centerline.

The computations shown in this study are based on videos of a representative cell which originally beats at $\sim$50 Hz. The cell is fully synchronized by flows along different directions ($\theta=0\degree,\ \pm45\degree$ and $90\degree$) at 49.2 Hz. In the computations, the applied flows are set to have an amplitude of 780 $\mu$m/s to reflect the experiments. Computations begin with the onset of the background flows (notified experimentally by a flashlight event), and last for $\sim$30 beats (500 frames sampled at 801 fps). Additionally, we confirm the results of \tflow-synchronization, that both flagella spend large fractions of time beating against the flows, with other cells and with \tflows at other frequencies. 

\subsection*{Isolate loads of external flows}
The total loads ($\textbf{F}$ and $P$) computed consist of two parts, one from the flow created by the two flagella themselves and the other from the applied flow. In the low Reynolds number regime, the loads of the two parts add up directly (linearity): $\textbf{F} = \textbf{F}_{\rm Self} + \textbf{F}_{\rm Flow}$, and $P= P_{\rm Self} + P_{\rm Flow}$. To isolate $\textbf{F}_{\rm Flow}$ and $P_{\rm Flow}$, we compute $\textbf{F}' = \textbf{F}_{\rm Self}$ and $P' = P_{\rm Self}$ by running the computation again but without the external flows, and obtain $\textbf{F}_{\rm Flow} = \textbf{F} - \textbf{F}'$ and $P_{\rm Flow} = P  - P'$. 

\subsection*{Modeling parameters}
We assume the flagellar intrinsic frequencies $f_c$ and $f_t$ to be 45~Hz and~65 Hz respectively\cite{Kamiya1987, Okita2005, Wan2013}. On this basis, $\lambda_{\rm c}:\lambda_{\rm t}$ is assumed to be 4:1 to account for the observed $f_0$ ($\sim50$ Hz). $\varepsilon_{\rm c}:\varepsilon_{\rm t}$ is set as 2:1, 1:1, and 1:2 for the \cflows, the \aflows, and the \tflows respectively, see \figref{fig:BEM_CATFlow}A-C. Additionally, $\varepsilon_{\rm c}+\varepsilon_{\rm t}$ is assumed to be constant to reflect the fact that $\mean{F}_{\rm Flow}^{\rm c} + \mean{F}_{\rm Flow}^{\rm t}$ approximately does not vary with flow directions. We take a typical value of $T^{\rm c,t}_{\rm eff}=1.57$~rad$^2$/s~\cite{Quaranta2015}. The sum of inter-flagellar coupling $\lambda_{\rm tot} = \lambda_{\rm c}+\lambda_{\rm t}$ is set to be large enough, i.e., $\lambda_{\rm tot}=3\nu_{\rm ct}$ with $\nu_{\rm ct}=|f_{\rm t}-f_{\rm c}|$, to account for the fact that: 1) the coordinated beating is approximated in-phase, and 2) up until the strongest flow applied, the coordinated beating cannot be broken (quantitative evaluation is detailed in \SSecModel). To model {\it wt} cells under calcium depletion, we decrease $\varepsilon_{\rm c}+\varepsilon_{\rm t}$ by 15\% - which is the mean decrease in the observed \ec,\ea, and \et (\figref{fig:ct_wt}E). For {\it ptx1} cells, we assume a symmetric inter-flagellar coupling ($\lambda_{\rm c}=\lambda_{\rm t}$) and a stronger noise level (\SSecPtx). The parameters are summarized in \tableref{table:parameters}.

\begin{table}[bhpt]
\centering
\caption{Modeling parameters}
\begin{tabular}{llrrr}
variable & symbol (unit) & TRIS & EGTA & \textit{ptx1}\\
Intrinsic freq.\cite{Kamiya1987,Okita2005} & $f_c,\ f_t$ (Hz) & 45,65 & 45,65 & 45,65\\
Basal coupling$^*$&$\lambda_c+\lambda_t$ (Hz) & 60 & 60 & 60\\ 
\textit{cis} dominance\cite{Okita2005,Horst1993}&$\lambda_c:\lambda_t$ (-) & 4:1 & 4:1 & 1:1\\
Flow detuning &$\nu$ (Hz) &[-10,10]&[-10,10]&[-10,10]\\
Total forcing\cite{Klindt2016}&$\varepsilon_{\rm c}+\varepsilon_{\rm t}$ (Hz) & 4.8 & 4.08 & 4.8\\
Noise$^*$\cite{Quaranta2015} &$T^{\rm c,t}_{\rm eff}$ (rad$^2$/s) & 1.57 & 1.57 & 9.42\\ 
\end{tabular}\label{table:parameters}
\vspace{3mm}
\\$*$ detailed in \SSecModel
\vspace{0mm}
\end{table}

\clearpage
\bibliography{reference}
\bibliographystyle{naturemag}

\section*{Acknowledgments}
The authors thank Roland Kieffer for technical support. D.W. thanks Ritsu Kamiya for helpful discussions. The authors acknowledge support by the European Research Council (ERC starting grants no. 716712 and no. 101042612). 

\section*{Author Contributions}
D.W. performed experiments, computations, designed the model, and drafted the manuscript. G.Q. performed early experiments and obtained preliminary results. M.A. and D.T. conceived the study, supervised the project and critically revised the manuscript.

\section*{Competing interests}
Authors declare that they have no competing interests.


\section*{Supplementary materials}
Supplementary Text\\
Figs. S1 to S5\\
References \cite{Polin2009,Quaranta2015,Wan2013,Kamiya1987,Kamiya2000,Okita2005,Horst1993,Leptos2013}\\

\clearpage

\clearpage

\clearpage

\clearpage

\clearpage

\end{document}


\baselineskip24pt

\maketitle 

\clearpage
\renewcommand{\theequation}{S\arabic{equation}}
\renewcommand{\thesection}{\textbf{S\arabic{section}}}
\renewcommand{\thefigure}{S\arabic{figure}}

\section{Extracting coupling strength by fitting phase dynamics}
In the work described in the manuscript, the flagellum-flow coupling strength $\varepsilon$ in {\it wt} cells is mainly extracted by the synchronization profile $\tau(\nu)\geq$50\%. Meanwhile, in previous works \cite{Polin2009,Quaranta2015}, fitting the distribution of phase dynamics is employed to extract $\varepsilon$. In the latter approach, the idea is that the phase locking during synchronization leads to a peaked probability distribution of $\Delta\varphi$, whose width is affected by the effective noise $T_{\rm eff}$. The distribution, $P(\Delta\varphi)$, can be derived from the Adler equation {\bf Eq.\,(1)} as:
\begin{equation}
    P(\Delta\varphi) = \int_{\delta_{\rm ct}}^{\Delta\varphi+2\pi} \exp(\frac{V(\Delta\varphi')-V(\Delta\varphi)}{T_{\rm eff}}) d\Delta\varphi'.
\label{eqPD}
\end{equation}
Here $V(\Delta\varphi) = \nu \Delta \varphi + \varepsilon \cos(\Delta\varphi)$ is a wash-board potential, $T_{\rm eff}$ is the noise, and $\Delta\varphi$ is the difference between the flagellar phase and the flow's phase.

\begin{figure}[hbt]
    \centering
    \includegraphics[width=0.6\textwidth, keepaspectratio=true]
        {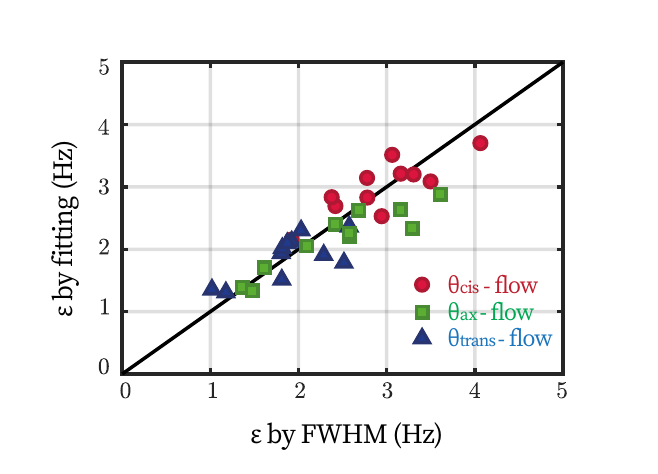}
    \vspace{-3mm}
    \caption{{\bf Equivalence of extracting coupling strength $\varepsilon$ by different methods.} Each point represents one cell under either the \aflow (green square), the \cflow (red circle), or the \tflow (blue triangle). The $x$ coordinate is the coupling strength $\varepsilon$ measured by the half width of synchronization profile $\tau(\nu) \geq 50\%$; and the y coordinate is obtained by fitting the flagellar phase dynamics.}
    \label{twoMethods}
    \vspace{3mm}
\end{figure}

Here, we demonstrate that these two approaches are equivalent in extracting $\varepsilon$. For all {\it wt} cells tested in the TRIS-minimal medium (N=11), their $\varepsilon(\theta)$ measured by the $\tau(\nu)$ width and extracted from fitting are plotted against each other, \figref{twoMethods}. All points center around the identity line, showing the equivalence in obtaining $\varepsilon$ by the two methods. For the {\it ptx1} dataset, $\varepsilon$ are extracted from fitting the phase dynamics.

\clearpage

\section{Hydrodynamic computation for flow along 90 degree}

\begin{figure}[htb!]
\begin{center}
\includegraphics[width=0.5\textwidth, keepaspectratio=true]
      {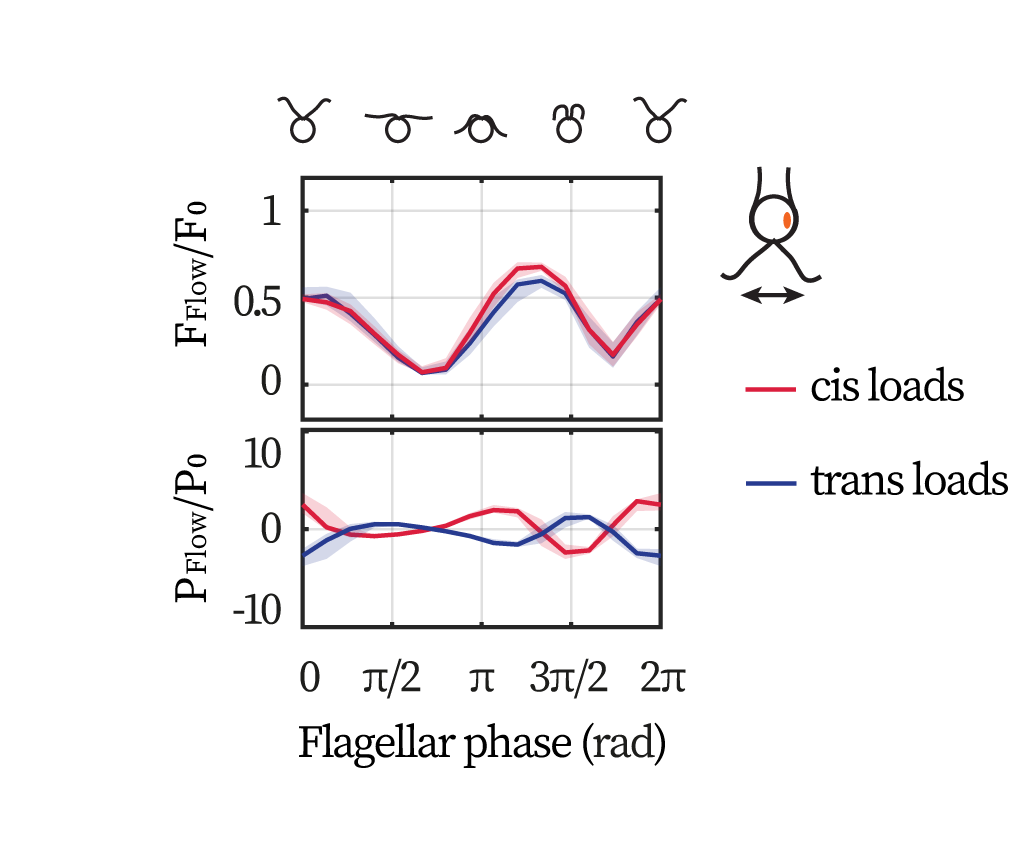}
      \end{center}
      \vspace{-3mm}
      \caption{{\bf Computed hydrodynamic loads on the flagella.} Computation results of the drag force (upper panel) and the force's rate of work (lower panel) on the {\it cis} (red) and the {\it trans} (blue) flagellum during synchronized cycles, when the cell is subjected to the flow with $\theta=90\degree$. Scaling factors $F_0$=9.9 pN and $P_0$=1.1 fW.}
      \label{fig:BEM_dyna}
\end{figure}

Similar to {\bf Fig.~2} in the main text, we present the computed drag force and power for the flow along $90\degree$. 
The solid lines and the shadings represent the median and the interquartile range of $F_{\rm Flow}$ and $P_{\rm Flow}$ over the flow-synchronized beats, respectively. Force magnitudes are scaled by $F_0 = 6\pi\mu R U_0$ = 9.9 pN, which is the Stokes drag on a typical free-swimming cell (radius R = 5 $\mu$m, swim velocity $U_0$ = 110 $\mu$m/s); while the viscous powers are scaled by $P_0 = F_0 U_0= 6\pi\mu R U_0^2$ = 1.1 fW. Here $\mu$ = 0.95 mPa$\cdot$s is the dynamic viscosity of water at 22 $^o$C. Quantitatively, the mean force is 0.37$F0$ and 0.34$F0$ (\figref{fig:BEM_dyna} top panel) while the mean power is -0.2$P_0$ and -0.4$P_0$ (\figref{fig:BEM_dyna} bottom panel), for the \textit{cis} and the \textit{trans} respectively.

\clearpage

\section{The model}
The external flow and the two flagella are described by three coupled ordinary differential equations (ODEs). Phase dynamics of these equations are examined by Monte-Carlo simulation. The temporal resolution of simulation ($dt$) is 1 ms, which corresponds to the experimental frame rates (801 Hz). 

\begin{subnumcases}\\
    \frac{d \varphi_{\rm f}}{dt} = 2 \pi f_{\rm f} \label{eq:3Body_flow}\\
    \frac{d \varphi_{\rm c}}{dt} = 2 \pi f_{\rm c} 
                - 2 \pi \lambda_{\rm t} \sin(\varphi_{\rm c} - \varphi_{\rm t}) 
                - 2 \pi \varepsilon_{\rm c} \sin(\varphi_{\rm c} - \varphi_{\rm f})
                + \zeta_{\rm c}(t) \label{eq:3Body_cis}\\
    \frac{d \varphi_{\rm t}}{dt} = 2 \pi f_{\rm t} 
                - 2 \pi \lambda_{\rm c} \sin(\varphi_{\rm t} - \varphi_{\rm c}) 
                - 2 \pi \varepsilon_{\rm t} \sin(\varphi_{\rm t} - \varphi_{\rm f})
                + \zeta_{\rm t}(t) \label{eq:3Body_trans}.
\end{subnumcases}

The {\it cis}, the {\it trans}, and the external flow are described as oscillators, whose intrinsic frequencies are $f_{\rm c,t,f}$ and phases $\varphi_{\rm c,t,f}$, respectively. The flow is assumed to be noise free and the two flagella are assumed to have the same level of noise ($\zeta_{\rm c}=\zeta_{\rm t}$). The noises are assumed to be Gaussian, $\langle \zeta_{\rm c,t} (\tau + t) \zeta_{\rm c,t}(\tau) \rangle$ = 2$T^{\rm c,t}_{\rm eff} \delta(t)$.

\subsection{Flagellar synchronization}
Setting $\varepsilon_{\rm c}$ and $\varepsilon_{\rm t}$ to 0, the interaction between the two flagella in the absence of the flow is modeled by:
\begin{subnumcases}\\
    \frac{d \varphi_{\rm t}}{dt} = 2 \pi f_{\rm c} 
                - 2 \pi \lambda_{\rm t} \sin(\varphi_{\rm c} - \varphi_{\rm t})
                + \zeta_{\rm c}(t) \label{eq:2Body_cis}\\
    \frac{d \varphi_{\rm c}}{dt} = 2 \pi f_{\rm t} 
                - 2 \pi \lambda_{\rm c} \sin(\varphi_{\rm t} - \varphi_{\rm c})
                + \zeta_{\rm t}(t) \label{eq:2Body_trans}.
\end{subnumcases}

When the two flagella are able to beat synchronously, $\frac{d \varphi_{\rm c}}{dt}=\frac{d \varphi_{\rm t}}{dt}=f_0$, we can obtain the analytical expression of $f_0$ by adding up $\lambda_{\rm c} \times$\eqref{eq:2Body_cis} and $\lambda_{\rm t} \times$\eqref{eq:2Body_trans}:
\begin{equation}
     f_0 = \frac{ \lambda_{\rm t}f_{\rm t}+\lambda_{\rm c}f_{\rm c}}{\lambda_{\rm c}+\lambda_{\rm t}}.
\end{equation}
Meanwhile, the steady-state phase difference $\delta_{\rm ct}=\varphi_{\rm c} - \varphi_{\rm t}$ is obtained by subtracting \eqref{eq:2Body_cis} from \eqref{eq:2Body_trans}:
\begin{equation}
    \sin(\delta_{\rm ct})=\frac{f_c-f_t}{\lambda_{\rm c}+\lambda_{\rm t}}=\frac{\nu_{\rm ct}}{\lambda_{\rm tot}}.
\end{equation}
It is therefore obvious that the two flagella can only beat at the same frequency ($d \varphi_{\rm c}/dt=d \varphi_{\rm t}/dt=f_0$) if $|\nu_{\rm ct}/\lambda_{\rm tot}|\leq1$. 

\subsection{Interaction between three oscillators}
Now we put the flow back into the picture. According to experimental observations, the two flagella mostly beat synchronously, we therefore focus on this case and first simplify the equations. By adding up $\lambda_{\rm c} \times$\eqref{eq:3Body_cis} and $\lambda_{\rm t} \times$\eqref{eq:3Body_trans}, and substituting $\varphi_{\rm c,t}$ as $\varphi_{\rm 0} = \varphi_{\rm c} - \delta_{\rm ct}/2 =\varphi_{\rm t} + \delta_{\rm ct}/2$, we obtain:
\begin{equation}\label{eq:3Body_interaction_raw}
    \frac{d \varphi_0}{dt} = 2 \pi f_{\rm 0}
                          - 2 \pi \frac{\lambda_{\rm c} \varepsilon_{\rm c}}{\lambda_{\rm c} + \lambda_{\rm t}} \sin\left(\varphi_{\rm 0} - \varphi_{\rm f} - \frac{\delta_{\rm ct}}{2}\right)
                          - 2 \pi \frac{\lambda_{\rm t} \varepsilon_{\rm t}}{\lambda_{\rm c} + \lambda_{\rm t}} \sin\left(\varphi_{\rm 0} - \varphi_{\rm f} + \frac{\delta_{\rm ct}}{2}\right)
                          +\frac{\lambda_{\rm t} \zeta_{\rm t} + \lambda_{\rm c} \zeta_{\rm c}}{\lambda_{\rm c}+\lambda_{\rm t}}.
\end{equation}

Given different choices of coupling constants ($\lambda_{\rm c,t},\varepsilon_{\rm c,t}$), this equation would generate complex phase dynamics - as we shall see in the following sections. We first limit the discussion to small $\delta_{\rm ct}$ - as it is observed in our experiment as well as in \cite{Wan2013}. The model's asymptotic behavior at $\delta_{\rm ct}\approx0$ is: 
\begin{equation}\label{3Body_interaction_asmyptotic}
    \frac{d \varphi_0}{dt} = 2 \pi f_{\rm 0} 
                          - 2 \pi \varepsilon \sin(\varphi_{\rm 0} 
                          - \varphi_{\rm f}) + \zeta_0(t),
\end{equation}
where
\begin{equation}\label{eq:CouplingWeightedAverage}
    \begin{aligned}
        f_0 = \frac{ \lambda_{\rm t}f_{\rm t}+\lambda_{\rm c}f_{\rm c}}{\varepsilon_{\rm     tc}+\lambda_{\rm t}},\ 
        \varepsilon = \frac{ \lambda_{\rm t} \varepsilon_{\rm t}+\lambda_{\rm c} \varepsilon_{\rm c}}{\lambda_{\rm c}+\lambda_{\rm t}},\ 
        \zeta_0 = \frac{ \lambda_{\rm t} \zeta_{\rm t}+\lambda_{\rm c} \zeta_{\rm c}}{\lambda_{\rm c}+\lambda_{\rm t}}.
    \end{aligned}
\end{equation}

\begin{figure}[ht]
\begin{center}
\includegraphics[width=90mm, keepaspectratio=true]
      {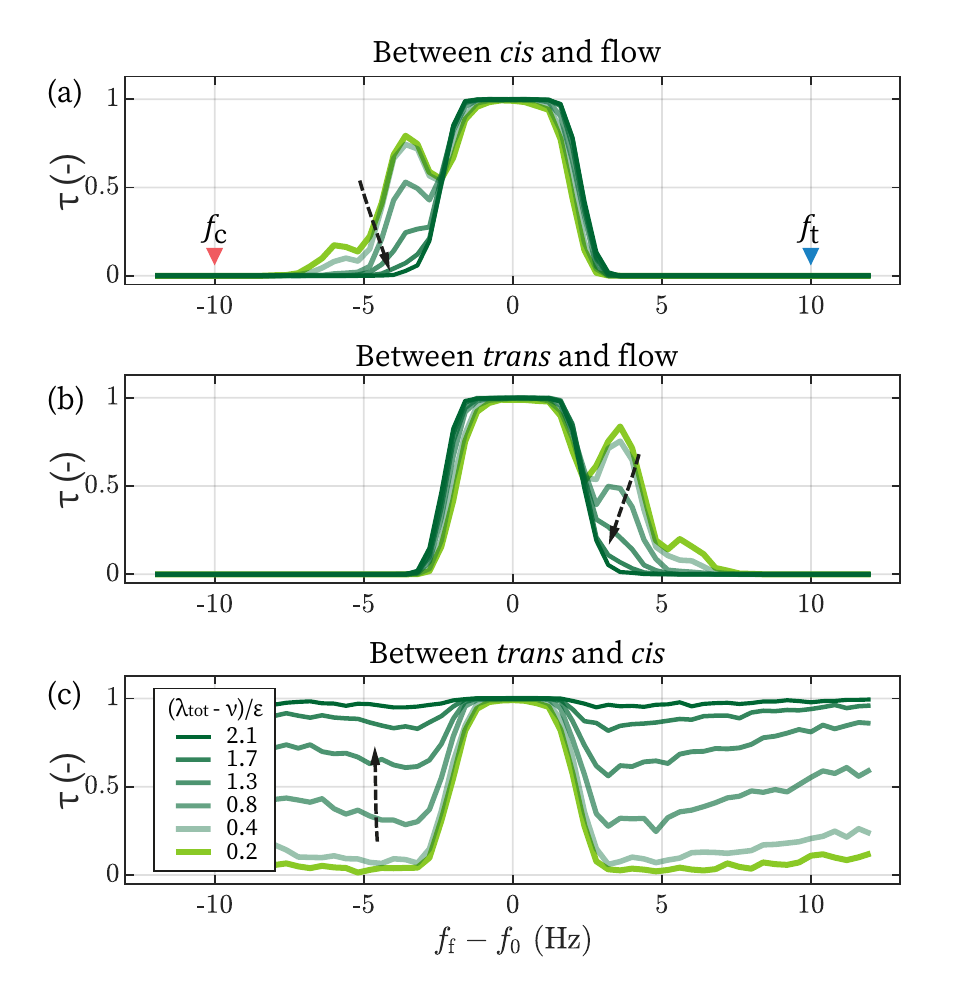}
      \end{center}
      \vspace{-5mm}
      \caption{{\bf Determine the lower limit of $\lambda_{\rm tot}$.} The time fractions of the {\it cis} (a) and the {\it trans} flagellum (b) synchronized by the flow. (c) The time fraction of where \textit{cis} and \textit{trans} are synchronized. Arrows points towards increasing $(\lambda_{\rm tot}-\nu)/\varepsilon$.}
      \label{fig:strongCoupling}
\end{figure}

In this strong-coupling limit ($\delta_{\rm ct}\approx0$, or equivalently, $\lambda_{\rm tot}\gg\nu_{\rm ct}$), the coupled flagella behaves as a single oscillator whose beating frequency $f_0$ will not be interfered by the external flow. The analytical form well captures the system's behavior, as shown by {\bf Fig.~5}F. Next we explore the model's behaviors when $\lambda_{\rm tot}-\nu_{\rm ct}$ is comparable with $\varepsilon$.

\subsection{Lower limit of inter-flagellar coupling}

The value $(\lambda_{\rm tot}-\nu_{\rm ct})/\varepsilon$ determines if the flow can disrupt the synchronization between \textit{cis} and \textit{trans}. We assume $\nu_{\rm ct}=20$ Hz\cite{Kamiya1987,Kamiya2000, Okita2005,Wan2013} and focus on synchronization of the \aflow. We plot the synchronization time fractions with increasing $\lambda_{\rm tot}$ in \figref{fig:strongCoupling}. When it satisfies $(\lambda_{\rm tot}-\nu_{\rm ct})/\varepsilon\geq2$, external flows cease to affect the flagellar synchronization observably. As the strongest flow ($21U_0$) applied experimentally corresponds to $\varepsilon\approx10$ Hz, altogether, we conclude that $\lambda_{\rm tot}\gtrsim\nu_{\rm ct}+2\varepsilon_{\rm max}=40$ Hz. In the main text, we set $\lambda_{\rm tot}=60=3\nu_{\rm ct}$ Hz, which satisfies this relation and matches the observation that the phase lag between the flagella ($\delta_{\rm ct}$) is small. 

\newpage

\clearpage

\section{Flagellar noise of the \emph{ptx1} mutant}
Here we show an as-yet uncharacterized strong noise present in the synchronous beating of the mutant {\it ptx1}. The in-phase (IP) mode of {\it ptx1} cells and the breaststroke beating of the {\it wt} cells are similar in waveform and frequency \cite{Horst1993,Leptos2013}. However, the former has a much stronger noise. 

\begin{figure}[hbt]
    \centering
    \includegraphics[width=130mm, keepaspectratio=true]
        {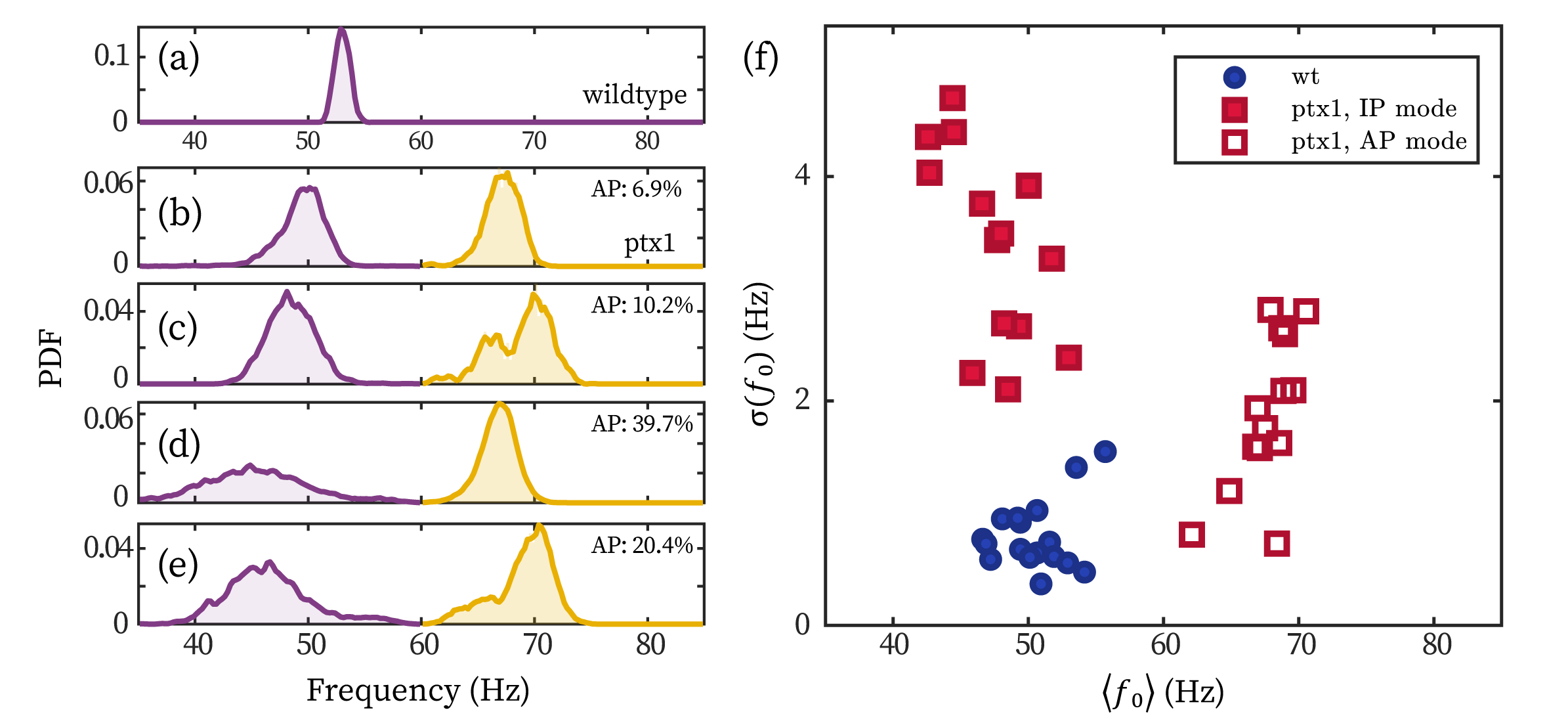}
    \vspace{-3mm}
    \caption{{\bf Stronger frequency fluctuation of the IP mode of \emph{ptx1} cells.} (a-e) Representative probability distributions of the beating frequency of a {\it wt} (a) and four {\it ptx1} cells (b-e) over 30 seconds. Probability distributions of the IP (purple) and AP mode (yellow) are respectively normalized for better visualization. The time fractions of the AP mode are noted in each panel. (f) The {\it wt} and {\it ptx1} cells represented by its mean beating frequency $\langle f_0 \rangle$ and the standard deviation of the beating frequencies over time $\sigma(f_0)$.}
    \label{fig:freqPDF_ptx1}
    \vspace{0mm}
\end{figure}

The strong noises show obviously in fluctuations of IP beating frequencies \cite{Leptos2013}.

In \figref{fig:freqPDF_ptx1}, we display the distribution of beating frequency of a representative {\it wt} cell (panel a) and four representative {\it ptx1} cells (panels b-e). The broad peaks of the IP (purple) and AP (yellow) beating of \textit{ptx1} sharply contrast the narrow peak of \textit{wt}. We quantify the frequency fluctuations of all the cells in the main text (N=11 for \textit{wt} and N=14 for \textit{ptx1}), \figref{fig:freqPDF_ptx1}f. The cells are represented by its mean beating frequency over time $\langle f_0 \rangle$ and the frequency's standard deviation $\sigma(f_0)$. Clearly, the breaststroke beating of {\it wt}, the IP, and the AP mode of {\it ptx1} each forms a cluster. The {\it wt} cluster is at $(\langle f_0 \rangle,\sigma(f_0)) = (50.5 \pm 2.6,\ 0.8 \pm 0.3)$ Hz (mean$\pm$ 1 std. the over cell population); and it is evidently less dispersed than both the IP and the AP mode of {\it ptx1}, which are at $(47.4 \pm 3.1,\ 3.4 \pm 0.9)$ Hz and $(67.6 \pm 2.1,\ 1.9 \pm 0.7)$ Hz, respectively. Under the assumption of a white (Gaussian) noise, $\sigma(f_0)$ is proportional to the noise level $\zeta$, and thus scales with $\sqrt{T_{\rm eff}}$. Consider that $\sigma(f_0)$ for {\it ptx1} is 3-5 folds larger than that of {\it wt}, we therefore conclude that the noise level in \textit{ptx1} is an order of magnitude larger than \textit{wt}, $T^{ptx1}_{\rm eff}/T^{wt}_{\rm eff}\sim\mathcal{O}(10)$.

\begin{figure}[hbt]
    \centering
    \includegraphics[width=110mm, keepaspectratio=true]
        {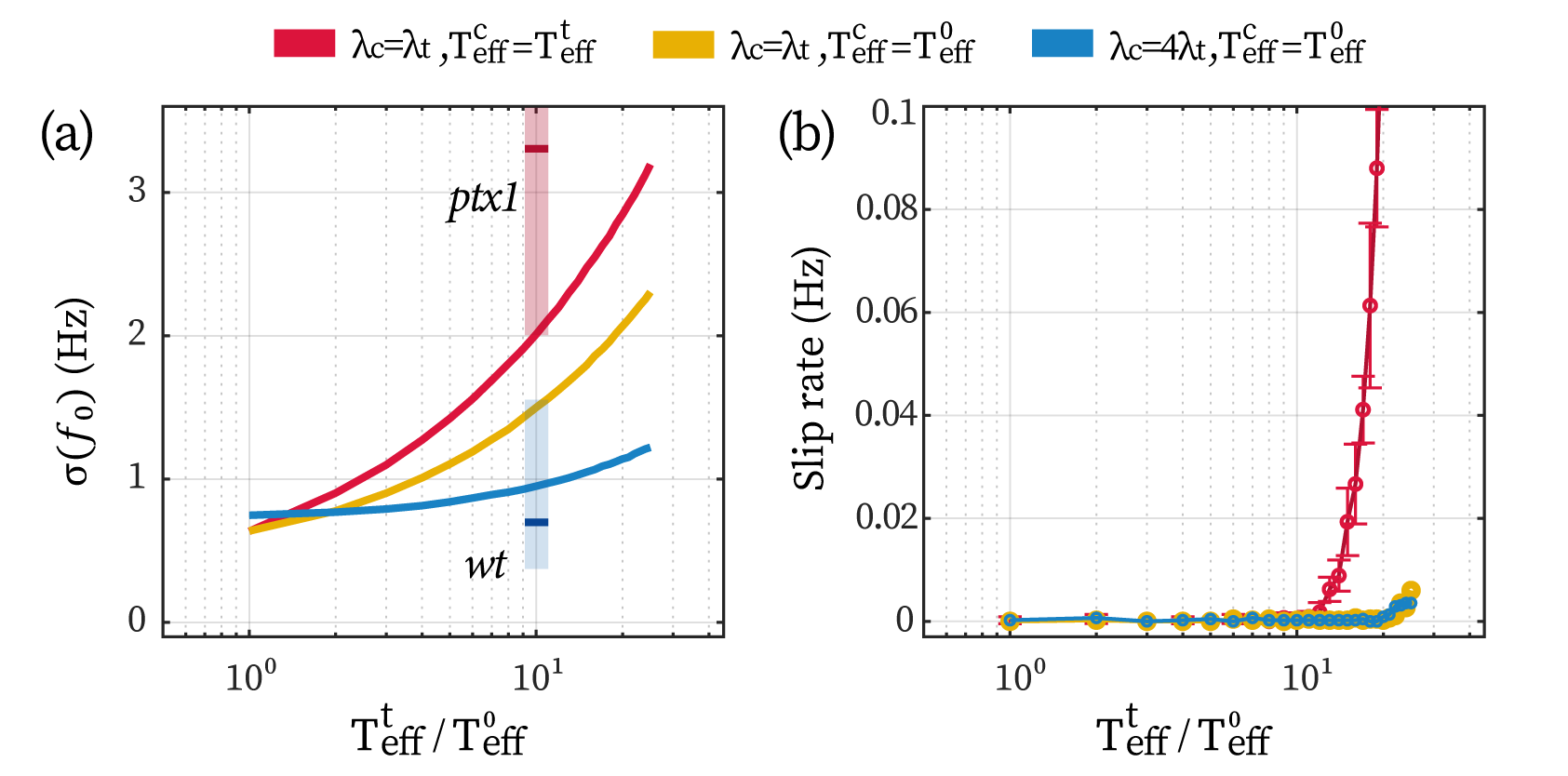}
    \vspace{-3mm}
    \caption{{\bf Effect of a low-noise \textit{cis} in stabilizing the beating of the \textit{trans}} (a) Fluctuations in beating frequency ($\sigma(f_0)$) under different coupling schemes and flagellar noises. Other model parameters are the same as used in the main text. The red and blue shaded area represent the experimentally observed range for \textit{ptx1} and \textit{wt} cells, respectively, with short bars marking the mean values. (b) the rate of slip under the conditions. Error bars correspond to 1 std. over N=9 repetitions.}
    \label{fig:noiseSuppression}
    \vspace{0mm}
\end{figure}

The stronger noise in \textit{ptx1} can be attributed to two sources, namely, the loss of a stable \textit{cis} and the loss of the unilateral coupling, \figref{fig:noiseSuppression}. We perform Monte-Carlo simulations of the coupled beating of \textit{cis} and \textit{trans} under three conditions: (1) a stable \textit{cis} ($T^{c}_{\rm eff}=T^0_{\rm eff}=1.57\ {\rm rad}/s^2$) coupled with the \textit{trans} unilaterally ($\lambda_{c} = 4\lambda_{t}$), (2) a stable \textit{cis} coupled with the \textit{trans} bilaterally ($\lambda_{c} = \lambda_{t}$), and (3) an equally noisy \textit{cis} ($T^{c}_{\rm eff}=T^t_{\rm eff}$) bilaterally coupled with \textit{trans}, see the blue, yellow, and red data in \figref{fig:noiseSuppression} respectively. It is obvious that, when the \textit{trans} is coupled to a stable \textit{cis}, varying its noise over an order of magnitude only leads to a $\sim20\%$ stronger frequency fluctuation (the blue line in \figref{fig:noiseSuppression}(a)). On the contrary, lacking either the unilateral coupling or the low-noised \textit{cis} would increase the fluctuation for 200\% (yellow line) or 300\% (red line). Qualitatively, simulation results are in agreement with experimental measurements assuming that $T^t_{\rm eff}/T^{c}_{\rm eff}\sim\mathcal{O}(10)$, see the red and blue shaded areas in \figref{fig:noiseSuppression}(a). Moreover, a low-noise \textit{cis} is already sufficient to prevent slips from interrupting the synchrony between \textit{cis} and \textit{trans}, even for bilateral coupling. In \figref{fig:noiseSuppression}(b), as long as the \textit{cis}-noise remains low, slips will be sparse ($<0.01$~Hz). Together, these simulation results highlight the stabilizing effect of a low-noise \textit{cis} flagellum, and illustrates the contribution of unilateral coupling in further enhancing the stabilization.

\bibliographystyle{naturemag}
\bibliography{reference}